\pdfoutput=1 
\documentclass{JINST}
\let\ifpdf\relax

\usepackage{graphicx}
\usepackage{caption}
\usepackage{subcaption}
\usepackage{amsmath}

\usepackage{appendix}

\newcommand{\muinj}{\ensuremath{\mathrm{\mu_{I}}}}

\newcommand{\sigmainj}{\ensuremath{\mathrm{\sigma_{I}}}}

\newcommand{\GD}{\ensuremath{\mathrm{G_{D}}}}
\newcommand{\GT}{\ensuremath{\mathrm{G_{T}}}}
\newcommand{\Ginj}{\ensuremath{\mathrm{G_{I}}}}
\newcommand{\muD}{\ensuremath{\mathrm{\mu_{D}}}}
\newcommand{\muT}{\ensuremath{\mathrm{\mu_{T}}}}
\newcommand{\sigmaD}{\ensuremath{\mathrm{\sigma_{D}}}}
\newcommand{\sigmaT}{\ensuremath{\mathrm{\sigma_{T}}}}

\newcommand{\npe}{\ensuremath{\mathrm{N_{\gamma e}}}}

\newcommand{\NHinj}{\ensuremath{\mathrm{N^{H}_{I}}}}
\newcommand{\GHinj}{\ensuremath{\mathrm{G^{H}_{I}}}}
\newcommand{\muHinj}{\ensuremath{\mathrm{\mu^{H}_{I}}}}
\newcommand{\sigmaHinj}{\ensuremath{\mathrm{\sigma^{H}_{I}}}}
\newcommand{\NHD}{\ensuremath{\mathrm{N^{H}_{D}}}}
\newcommand{\GHD}{\ensuremath{\mathrm{G^{H}_{D}}}}
\newcommand{\muHD}{\ensuremath{\mathrm{\mu^{H}_{D}}}}
\newcommand{\sigmaHD}{\ensuremath{\mathrm{\sigma^{H}_{D}}}}

\newcommand{\NT}{\ensuremath{\mathrm{N_{t}}}}
\newcommand{\NB}{\ensuremath{\mathrm{N_{b}}}}
\newcommand{\NS}{\ensuremath{\mathrm{N_{s}}}}
\newcommand{\NE}{\ensuremath{\mathrm{N_{event}}}}

\newcommand{\Etransfer}{\ensuremath{\mathrm{E_{transfer}}}}
\newcommand{\Edrift}{\ensuremath{\mathrm{E_{drift}}}}
\newcommand{\Einduction}{\ensuremath{\mathrm{E_{induction}}}}

\newcommand{\mos}{\ensuremath{\mathrm{\frac{M\Omega}{square}}}}
\newcommand{\nech}{\ensuremath{\mathrm{Ne/(5\%)CH_4}}}
\newcommand{\kvcm}{\ensuremath{\mathrm{\frac{kV}{cm}}}}
\newcommand{\kev}{\ensuremath{\mathrm{KeV}}}

\newcommand{\effgain}{\ensuremath{\mathrm{2.5 \times 10^3}}}

\title{A concept for laboratory studies of radiation detectors over a broad dynamic-range: instabilities evaluation in THGEM-structures}

\author{S. Bressler\thanks{Corresponding
author.}, L. Moleri, L. Arazi, E. Erdal, A. Rubin, M. Pitt ~and A. Breskin\\
\llap{}Department of Particle Physics, Weizmann Institute of science,\\
  76100 Rehovot, Israel\\
  E-mail: \email{shikma.bressler@weizmann.ac.il}}

\abstract{A simple methodology for evaluating the dynamic-range of gas avalanche detectors in the laboratory is presented and discussed. It comprises two tools: a charge injector of tunable gain which transfers radiation-induced amplified electron swarms to the investigated detector to mimic events with well defined primary-ionization spectra, and a systematic approach for measuring the detector's discharge probability. The methodology, applicable to a broad range of detectors, is applied here for instability studies in various single-stage THGEM and THGEM-WELL structures. The results indicate upon a somewhat larger attainable dynamic range in a single-stage THGEM operated with additional multiplication in the induction gap.}


\begin{document}

\section{Introduction}

Most gas-avalanche radiation detectors, optimized for operation in a given radiation field, suffer from instabilities under highly ionizing background. This problem is of particular concern in modern Micro Pattern Gas Detectors (MPGDs), which - in spite of their higher rate capability and spatial resolution - are generally more limited in their dynamic range than traditional wire chambers. Considerable efforts are thus made in order to understand and minimize the occurrence and effects of discharges in MPGDs~\cite{Fonte2001,Fonte1997,Peskov10}. In this work, a concept for laboratory studies of the behaviour of gaseous detectors over a broad dynamic range is suggested and investigated; it permits exposing the detector simultaneously to selectable low and (tuned) high radiation fields - mimicking operation under a background of Highly Ionizing Particles (HIPs). The method is applied here to various detector configurations incorporating Thick Gas Electron Multiplier (THGEM)~\cite{Breskin09} elements.

Similarly to other MPGDs, the effective gas-gain of a THGEM increases exponentially with the applied voltage and does not reach saturation. At a given gain, the average avalanche charge increases linearly with the number of primary deposited electrons; discharges occur for avalanche sizes exceeding the Raether limit~\cite{Raether}, namely $\mathrm{10^6 - 10^7}$~- electrons -  resulting in detection dead time, efficiency loss and potential damage to the detector electrodes and its readout electronics.

The discharge limits in various gaseous detectors, including THGEMs, were investigated in different works, e.g. in~\cite{Fonte2001,Fonte1997,Peskov10,Bressan99, Bachmann02, Charles11}; they define an upper bound to the detector's dynamic-range. Theoretically, discharges occur when the number of radiation-induced primary electrons (PEs) multiplied by the gas-gain exceeds the Raether limit. However, this simplistic description is not always consistent with experimental observations. For example, it was shown recently that different THGEM structures, e.g. standard THGEM with induction gap and Thick WELL-structures (a THGEM closed at its bottom with metallic or resistive anodes~\cite{Arazi12}), can have substantially different discharge probabilities when operated at similar effective-gain values and exposed to the same type of radiation.

The response of a detector to HIPs should best be studied in dedicated irradiation facilities; e.g., minimum ionizing particle (MIP) detectors are often investigated with hadronic beams. The latter, however, are not easily accessible and usually provide limited broad radiation spectra. A method mimicking controlled HIP spectra would facilitate detector R\&D in the laboratory over a broad dynamic range - prior to further studies in accelerator facilities.

Two methodologies were developed in this work in this respect: the so-called "charge-injector method" - mimicking highly ionizing events, and a method to systematically measure the detector's discharge probability. They can be used independently, but when combined, they provide an efficient means for evaluating the detector's response to HIPs - in terms of its discharge probability as a function of the deposited primary charge.

Our present studies of discharge probabilities in various THGEM (standard and WELL) configurations were performed as part of a broader effort to develop very thin sampling elements for Digital Hadronic Calorimetry (DHCAL) in future linear colliders (ILC or CLIC~\cite{Brau13,Linssen13}), e.g., that of the Silicon Detector (SiD)~\cite{sid09}. Our accelerator-beam studies~\cite{Arazi12,Bressler13,Arazi13}~focused on thin (below $\sim$10 mm) sampling structures capable of operation in hadronic beams at gains of a few thousands; the structures considered had been previously evaluated with respect to their ability to meet the requirements of high detection efficiency, low pad multiplicity (digital pad counting) and small overall thickness. The results of our recent laboratory studies of some basic properties of the different THGEM configurations were reported in~\cite{Arazi13_2}.

The concept of the charge-injector as a tool to mimic HIPs in the laboratory, is introduced in section~\ref{sec:InjectorConcept}. The various experimental setups and methods are described in section~\ref{sec:Setup}. In section~\ref{sec:Validation} we discuss the conditions for an effective application of the charge-injection and discharge probability evaluation methods. The results of a comparative study of instabilities in various THGEM configurations, over a broad dynamic range are presented in section~\ref{sec:Results}. Section~\ref{sec:Conclusion} concludes the work and provides a discussion of the results. Simulation details are provided in the Appendix.

\section{The concept of a charge-injector}
\label{sec:InjectorConcept}

\begin{figure}[h]
  \centering
  \includegraphics[scale=1]{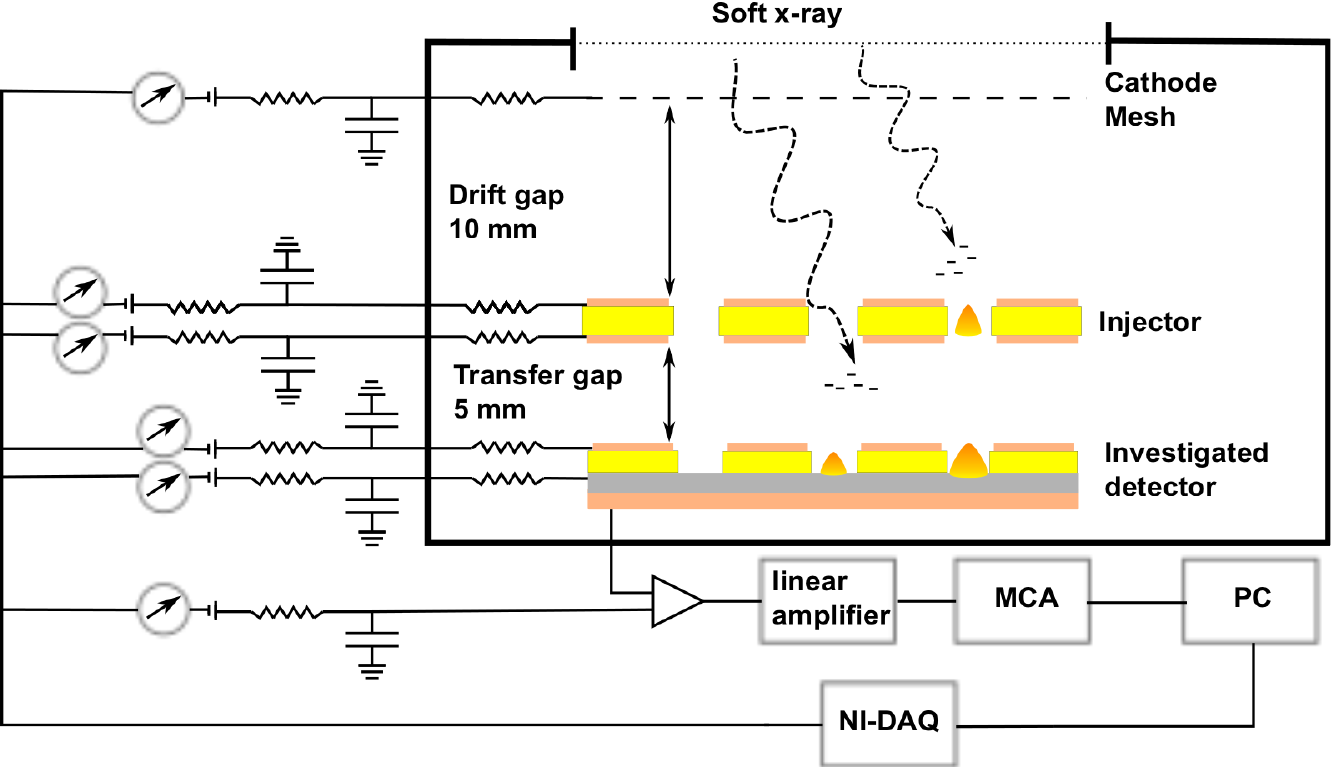}
  \caption{Schematic description of the charge-injector method. An investigated detector (here a RWELL) is preceded by a charge pre-amplifier, "injector" (here a THGEM). Impinging radiation (here soft X-rays) is converted within both, a conversion/drift gap and a transfer gap between the two elements. X-rays converted in the drift gap above the injector are pre-amplified; the resulting avalanche electrons are transferred into the investigated detector, through a transfer gap; the resulting avalanche mimics highly-ionizing events. Events converted within the transfer gap undergo charge multiplication only in the detector - mimicking low-ionization events.}
  \label{fig:chamber}
\end{figure}

As illustrated in figure~\ref{fig:chamber}, the charge-injector setup consists of a radiation-conversion and drift gap followed by a charge multiplier ("injector", here a THGEM) - preceding the investigated detector. Operating as a pre-amplification stage, the injector multiplies the ionization electrons released by the incoming radiation; the multiplied electron swarm, whose size is selected by tuning the THGEM pre-amplification (operating voltage), is transferred through a transfer gap to the investigated detector (for example, a Resistive WELL, RWELL~\cite{Arazi12} in figure~\ref{fig:chamber}).

By varying the injector gain (typically up to 100) it is possible to expose the investigated detector to a broad spectrum of primary ionization. In this work, the term "primary electrons" (PEs) refers to the electrons that underwent pre-amplification in the injector. Note that ionization electrons originating from X-ray conversion events occurring within the transfer gap (figure~\ref{fig:chamber}) are solely multiplied in the investigated detector.

It should be stressed that the difference between a charge-injector and a multi-stage Micro Pattern Gaseous Detector (MPGD) configuration is conceptual rather than technical. In a multi- stage MPGD, all the stages are considered as a single detector, characterized and optimized as a whole. In a setup containing a detector and an injector (figure~\ref{fig:chamber}), the two are assumed independent, and the injector is used as a tool to evaluate the detector. Therefore, it is important to define the conditions under which the response of the detector is not affected by the presence of the injector. These conditions are defined and discussed in section~\ref{sec:Validation}.

\section{Experimental setup and methodology}
\label{sec:Setup}

The experimental setup comprised an Oxford Series 5000 packaged X-ray tube with a copper target, a chamber containing the charge-injector and the investigated detector assembly (figure~\ref{fig:chamber}) and a readout system.

\begin{figure}[h]
  \centering
  \includegraphics[scale=0.5]{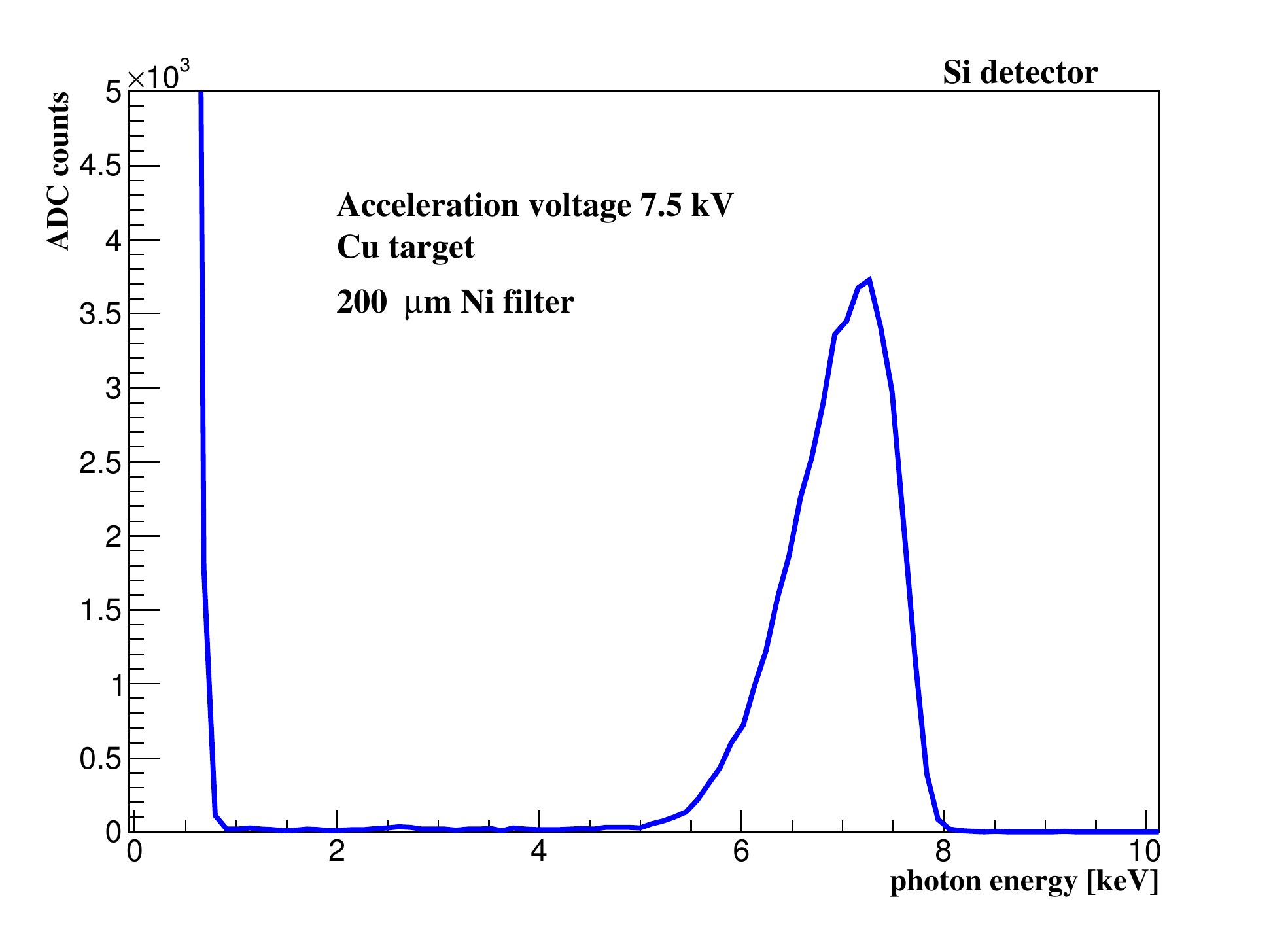} 
  \caption{The X-ray spectrum used throughout the experiments, measured with an Amptek XR100-CR silicon detector.}
  \label{fig:bremsstrahlung}
\end{figure}

To avoid eventual discharges originating from events in the high-energy Bremsstrahlung tail of the X-ray spectrum, the X-ray tube was operated as follows. The acceleration voltage was set to 7.5 kV, below the copper $\mathrm{K_{\alpha}}$~shell ionization energy; using a 0.2 mm thick Ni filter, the resulting spectrum, of low-energy Bremsstrahlung photons only, had a well-defined peak, as measured with an Amptek XR100-CR silicon detector (figure~\ref{fig:bremsstrahlung}). The slightly asymmetric distribution had a mean value of 6.7 \kev~(with FWHM of ~20\%) which, in a \nech~gas mixture, yielded on average 188 electron-ion pairs~\cite{Weiss57,Sauli77}.

The photon flux was collimated by an aluminium tube, keeping the event rate fixed at $\sim$100 Hz; the irradiated area on the detector was about 60 $\mathrm{mm^2}$. The X-ray photons penetrated into the detector chamber through a 25 $\mathrm{\mu m}$~thick Mylar window. The chamber was flushed with \nech~gas mixture at 1 atm, at a constant flow of 10 cc/min, maintained by an MKS 247 mass-flow controller.

The injector/detector configuration shown in figure~\ref{fig:chamber} (electrode size $\mathrm{30 \times 30~mm^2}$) had the following parameters. A cathode mesh was placed approximately 20 mm away from the Mylar window; the conversion/drift gap between the THGEM injector electrode and the cathode mesh was 10 mm; the transfer gap between the injector and the investigated detector was set at 5 mm (see section~\ref{sec:Validation}). The injector electrode was a double-sided 0.4 mm thick THGEM with 0.5 mm diameter holes drilled in an hexagonal pattern with 1 mm pitch, with 0.1 mm wide etched rims around the holes. The configurations of the different investigated detectors are described in section~\ref{sec:Results}.

The electrodes were biased with individual CAEN N471 High Voltage (HV) power supplies; all the HV channels were connected via 25 Hz low-pass filters. Apart from the anode, all the electrodes were connected via additional 10 $\mathrm{M\Omega}$~resistors (figure~\ref{fig:chamber}). The signals induced on the anode were recorded with an Ortec 124 charge-sensitive pre-amplifier followed by an Ortec 570 linear amplifier and an Amptek 8000A Multi Channel Analyzer (MCA). The MCA output was charge-calibrated.

\subsection{Distribution of the number of primary electrons}
\label{sec:methodInjector}

The injector allows for controlling the number of injected PEs and the event rate can also be controlled (by tuning the X-ray source operation current, addition of filters and collimation). The distribution of the number of PEs, namely the number of electrons after the injector stage, cannot be measured directly, due to its low operation gain. Therefore, the PE distributions, with their mean value ($\mathrm{\mu_{PE}}$) and width ($\mathrm{\sigma_{PE}}$), were extracted from the spectra measured on the investigated detector, as described below.

\begin{figure}[h]
  \centering
  \includegraphics[scale=0.5]{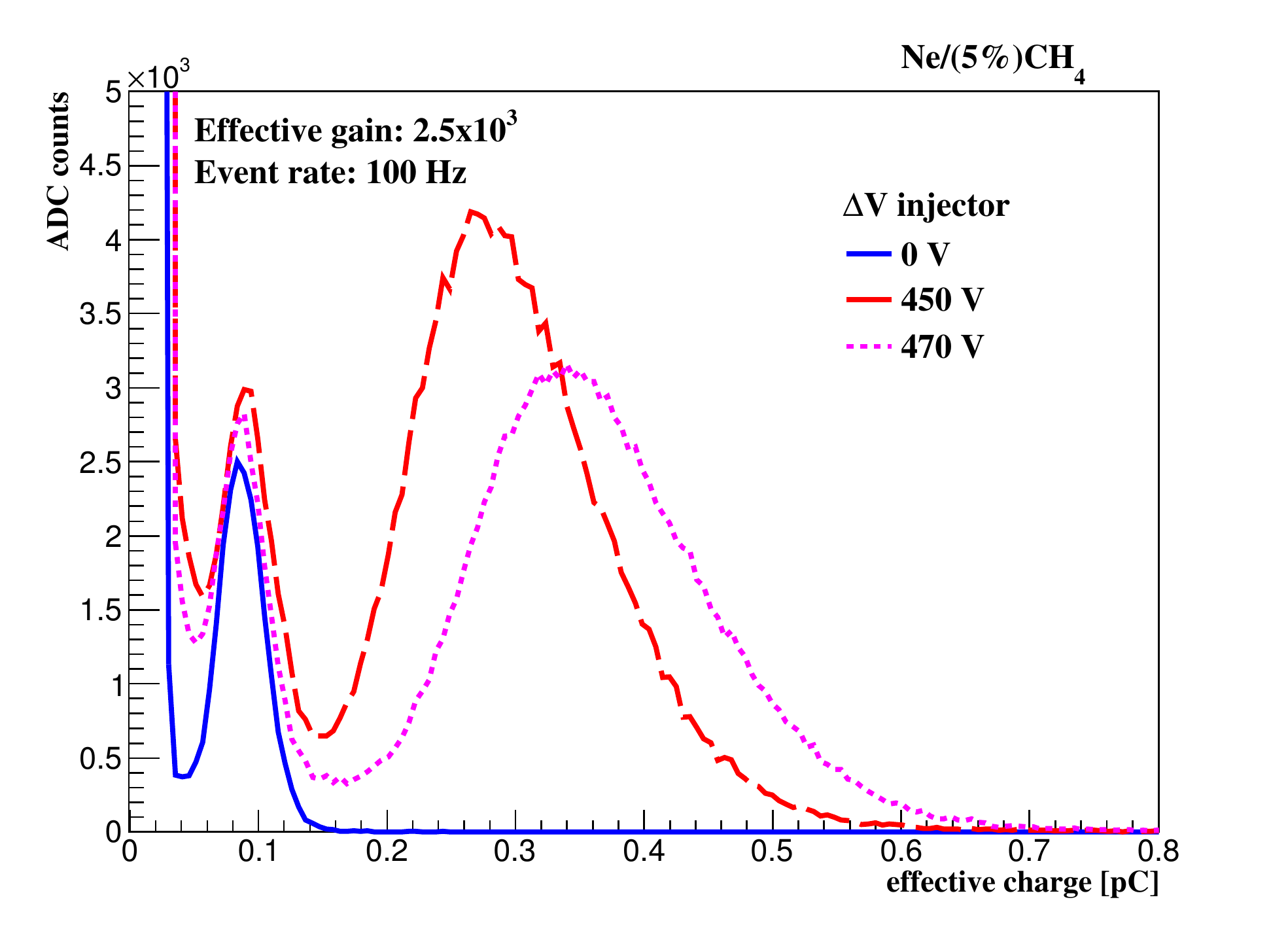}
  \caption{Typical spectra induced by the low-energy Bremsstrahlung photons (figure~\protect\ref{fig:bremsstrahlung}), recorded after multiplication in the injector/detector setup shown in figure~\protect\ref{fig:chamber} (here the investigated detector was a 0.4 mm double-sided THGEM, see text). The solid-line spectrum was measured when the injector was unbiased. The peak corresponds to photons converted within the transfer gap, with the electrons multiplied only in the investigated detector. The dashed- and dotted- spectra were obtained at injector-gains of $\sim$4 and $\sim$5 respectively, with low-charge peaks resulting from charges multiplied only in the detector and high-charge peaks resulting from multiplication in both the injector and the detector. The operation conditions, in \nech~gas mixture, are given in the text.}
  \label{fig:injectorSpectrum}
\end{figure}

Typical spectra of low-energy Bremsstrahlung photons, measured with a system comprising the injector and a detector, is shown in figure~\ref{fig:injectorSpectrum}. Here the investigated detector was 0.4 mm thick double-sided THGEM operated with 1 mm induction gap. The THGEM electrode had 0.5 mm hole-diameter, a pitch of 1 mm and 0.1 mm etched rims around the holes. The figure contains three spectra: one with a single low-charge peak value, $\mathrm{Q_{low}}$, corresponding to photons converted within the transfer gap between the injector and the detector, with the electrons multiplied only in the investigated detector; the injector was unbiased, \Etransfer~= 0.5 \kvcm, and the investigated-detector gain was \effgain. The two other spectra, with two distinct peaks (high and low), correspond to photons converted within the drift and transfer gaps, respectively; some of the resulting charges were multiplied in both the injector and the detector (higher peaks, $\mathrm{Q_{high}}$) and others were multiplied only in the detector (lower peaks, $\mathrm{Q_{low}}$); here \Edrift~= \Etransfer~= 0.5 \kvcm, the investigated-detector gain was \effgain~and the injector-gains were $\sim$4 and $\sim$5.

In what follows we assume that each X-ray photon converted in \nech~gas mixture on the average to \npe~= 188 electrons (6.7 \kev; average energy for electron-ion pair production w~35.8 eV~\cite{Weiss57,Sauli77}). We further assume a full electron transfer efficiency of the ionization electrons to the injector and / or the investigated-detector holes, under the applied field-values \Etransfer~= \Edrift~= 0.5 \kvcm~\cite{Shalem06}.

Typical X-ray spectra measured with a THGEM detector have Gaussian shapes (as indicated, for instance, by the low- and high-charge peaks of figure~\ref{fig:injectorSpectrum}). Therefore, we modeled the injector effective gain, that of the detector and the total effective gain - with three different Gaussian distributions: \Ginj(\muinj,\sigmainj), \GD(\muD,\sigmaD), and \GT(\muT,\sigmaT)~respectively. Here $\mathrm{G_X}$ stands for a Gaussian with a mean value $\mathrm{\mu_X}$~and a width $\mathrm{\sigma_X}$. \GD(\muD,\sigmaD)~and \GT(\muT,\sigmaT)~can be extracted directly from the measured spectra, by a Gaussian fit to the low-charge and high-charge peaks in the distributions, respectively.

Assuming that the distribution of the number of PEs is related to that of the injector gain through the constant term \npe~(in particular $\mathrm{\mu_{PE} = \npe \times  \muinj}$), characterizing the distribution of the number of PEs would mean estimating the values of \muinj~and \sigmainj. The latter is important since events in the higher part of the charge distributions have higher probability to cause a discharge.

\textbf{\linebreak Estimating \muinj:} We consider a detector operated at a mean gain \muD, an injector operated at a mean gain \muinj~and a source that induces on average \npe~converted electrons. Assuming 100\% electron transfer efficiency of photon-induced electrons in both the drift and transfer gaps, the mean value of the low-charge distribution, corresponding to electron multiplication only in the investigated detector, is given by $\mathrm{Q_{low} = \npe \times \muD}$. The mean value of the high-charge distribution, corresponding to multiplication in both the injector and the detector, is given by $\mathrm{Q_{high} = \npe \times \muinj \times \muD}$. 

With this definition of $\mathrm{Q_{high}}$, the extraction efficiency of avalanche electrons from the injector into the transfer gap and the collection efficiency of the fraction of these electrons into the detector were taken into account in the definition of \muinj. Therefore, \muinj~is estimated from the ratio between $\mathrm{Q_{high}}$~and $\mathrm{Q_{low}}$~values: $\mathrm{\muinj = \frac{Q_{high}}{Q_{low}}}$.

\textbf{\linebreak Estimating \sigmainj:} The width of the distribution of the injector gain (and the resulting number of PEs) was estimated from the toy Monte-Carlo simulations described in Appendix~\ref{appendixA}. It was shown that the width of the distribution of the injector gain is at most at the order of 15\%. This result is important as it shows that the distribution of the number of PEs is relatively narrow. In particular, it does not have any high-charge tail which, as mentioned earlier, would be likely to induce discharges at high detector gains.

\subsection{Evaluating discharge probability}
\label{sec:methodDischarge}

A discharge is defined here as a rapid increase in the current supplied to two or more detector electrodes. The current from the power supply was monitored with a National Instruments data acquisition (NI-DAQ) analogue signal digitization board NI-USB 6008. The signals were sampled at 10 Hz and recorded for further analysis using LabView Signal Express 2012~\cite{SignalExpress}.

\begin{figure}[h]
        
        \begin{subfigure}{0.5\textwidth}
               \includegraphics[scale=0.4]{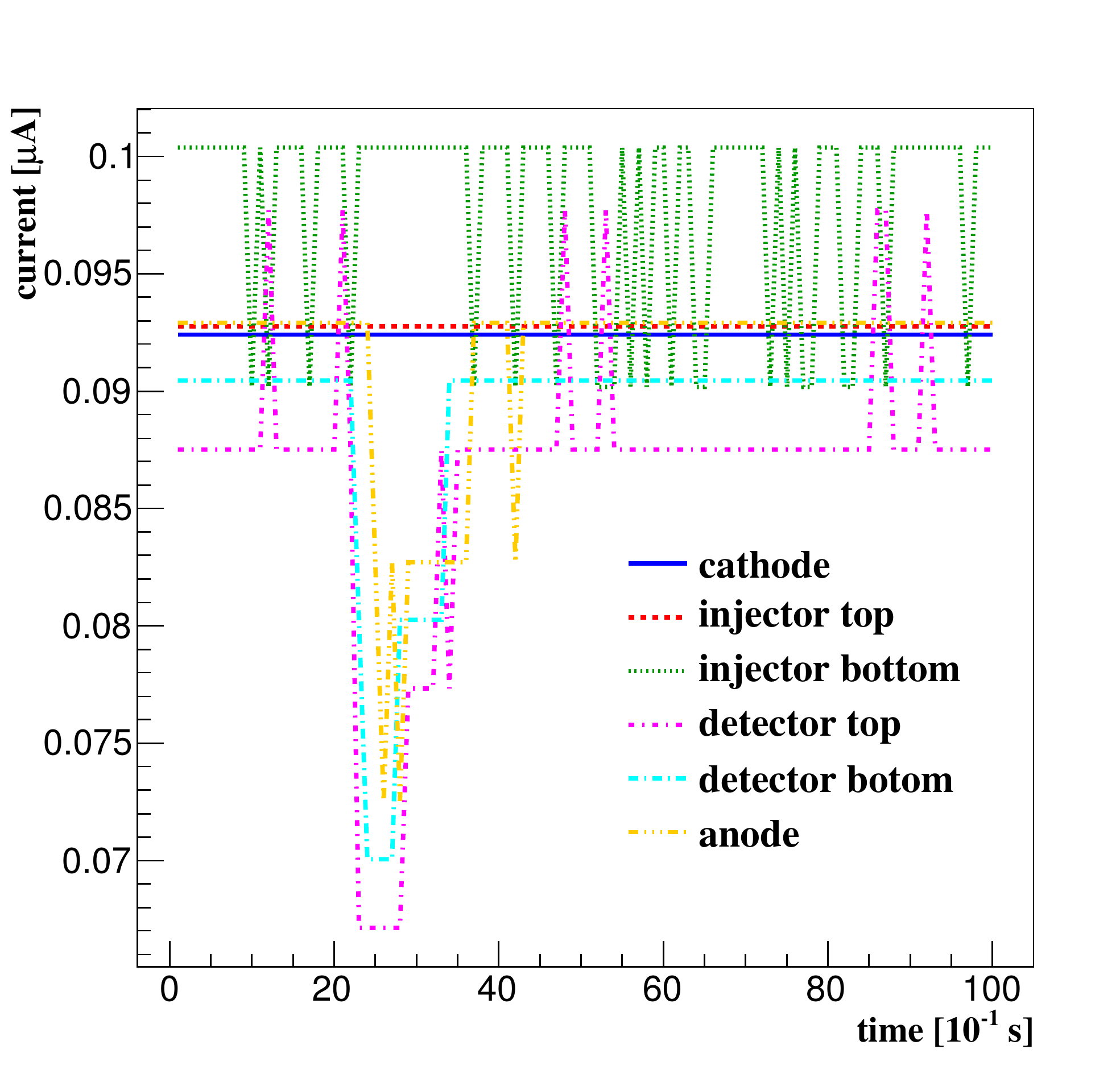}
               \hspace{.1\linewidth}\parbox[l]{.1\linewidth}\caption{}
                \label{fig:sparkTHGEM}
        \end{subfigure}%
        \quad 
        \begin{subfigure}{0.5\textwidth}
                
                \includegraphics[scale=0.4]{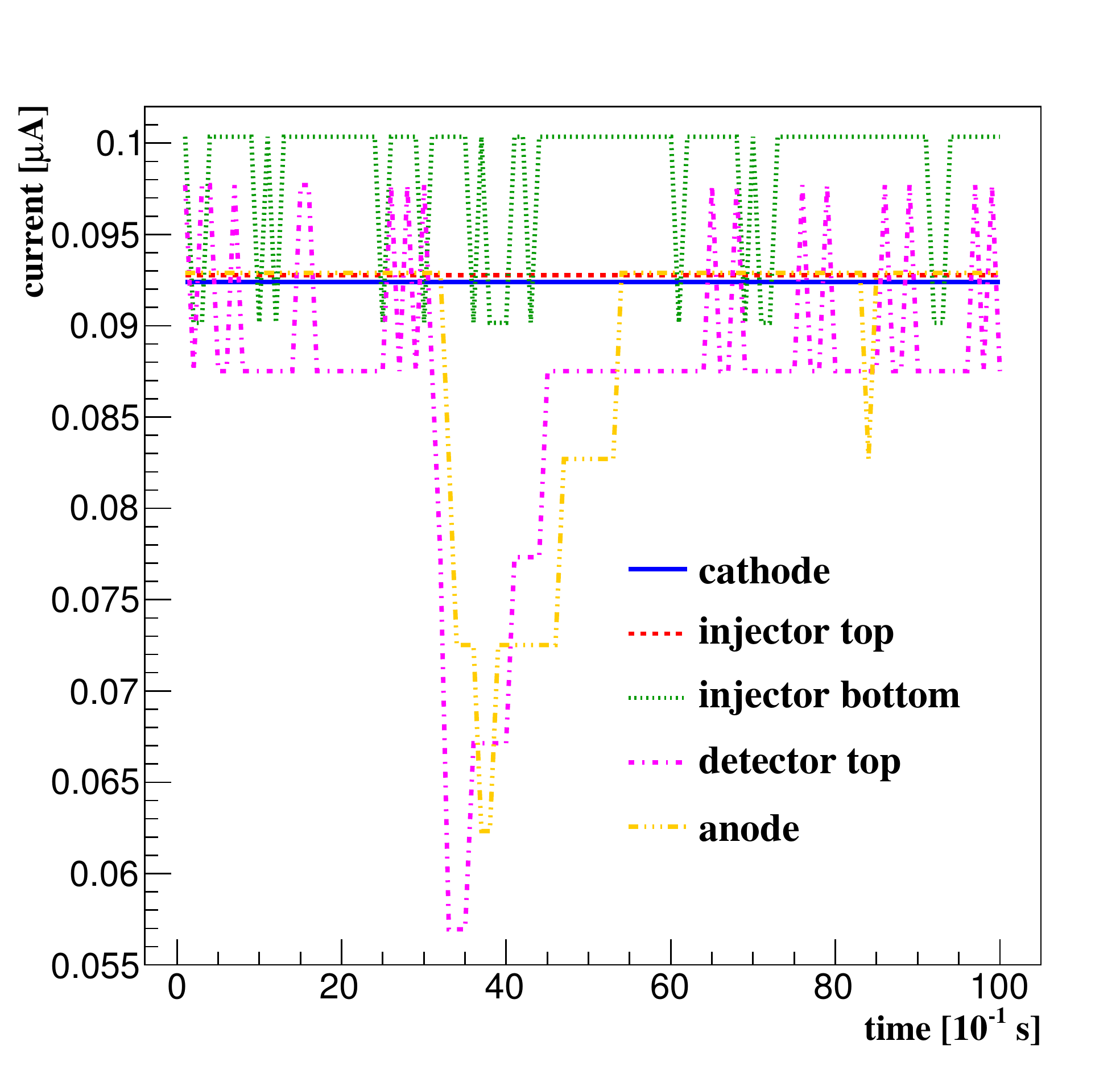}
\hspace{.1\linewidth}\parbox[l]{.1\linewidth}\caption{}
                \label{fig:sparkTHWELL}
        \end{subfigure}
        \caption{(a) A typical discharge in a single-THGEM detector with 1 mm induction gap (with multiplication in the THGEM and the gap). $\mathrm{\Delta V_{THGEM}}$~= 630 V, \Einduction~= 3 \kvcm, $\mathrm{\Delta V_{injector}}$~= 390 V. (b) A typical discharge in a THWELL detector, $\mathrm{\Delta V_{THWELL}}$~= 760 V, $\mathrm{\Delta V_{injector}}$~= 365 V.}
        \label{fig:discharges}
\end{figure}

Typical discharge measurements in a standard double-sided THGEM with an induction gap (with small multiplication in the gap) and in a single-sided Thick WELL (THWELL - a THGEM with a closed metal anode at its bottom~\cite{Arazi12}) are shown in figure~\ref{fig:discharges}. In both configurations, no current was supplied to the cathode and the injector's electrodes, indicating that the discharge and its origin are confined within the investigated-detector holes. In the double-sided THGEM configuration, both THGEM electrodes and the anode were affected by the discharge. The THWELL bottom-side is attached to the anode, hence the top electrode and the anode were affected by the discharge.

While we are interested in measuring the discharge probability of different THGEM-based detector structures as a function of the number of primary electrons, we noticed non-negligible rates of discharges occurring also without irradiation. These background discharges might have been induced by defects in the electrodes, cosmic events, natural radioactivity of the electrode material (i.e. glass-fibres of FR4), etc. A deeper on-going study of the contributions to source-less discharges is beyond the scope of this work.

The background discharges were handled as follows: we denote by \NS~the number of source-induced discharges, by \NB~the number of background discharges, by \NT~the total number of discharges (\NT~= \NS~+ \NB) and by \NE~the number of events (converted photons); the discharge probability is given by: $\mathrm{P = \frac{\NS}{\NE} = \frac{\NT - \NB}{\NE}}$

The rate of background discharges may depend on the detector configuration. Therefore, for each investigated structure, and for each HV configuration, two independent measurements were carried out: one of the total number of discharges, \NT, and another of the number of background discharges \NB~(the former with the X-ray source "on" and the latter with the X-ray source "off").

We excluded the first 10 seconds after every discharge to avoid possible influences of the detector and power-supply recovery times. Therefore, for a measurement of N discharges in T seconds, the effective measuring-time was defined as $\mathrm{T_{eff} = T - 10 \times N}$. A scaling factor was applied to \NB~to correct for differences in the duration of the two measurements; the number of background discharges with the source "on" was estimated as the number of discharges with the source "off" times the ratio between the effective times of the two measurements. Subtracting the number of background discharges from the total number of discharges, we assumed that the background discharges are Poisson-distributed, as justified in section~\ref{sec:Validation}. All the measurements were performed under similar conditions, with the same protocols, to reduce eventual systematic errors. 

By excluding the 10 seconds following a discharge from the measurements we limited the discharge rate measurement to a maximum rate of 0.1 discharges/sec; this rate is significantly higher than the typical measured background discharge rate (of order $\mathrm{10^{-3}}$~discharges/sec). Since the event rate was $\sim$100 events/sec, the maximum discharge probability that we were able to measure was $\mathrm{\sim 0.1/100 \simeq 10^{-3}}$~discharges/events.

It should be noted that the discharge probability of a THGEM-structure depends also on the electrodes' quality. In an attempt to make our measurements independent of the specific electrode that was used, we investigated only electrodes with low leakage currents ($\mathrm{\leq 5 \times 10^{-11} A}$~at 1 kV in air), that break down (suffer spontaneous discharges) at voltages not lower than those predicted by the Paschen law~\cite{Paschen1889}. In section~\ref{sec:Validation}~we justify this selection and show that detectors with such electrical properties have similar response to highly ionizing events.

Small shifts in the detector's and injector's gains were often observed along long (a couple of hours) measurements~\cite{Cortesi09}; Gain shifts up to 20\% in the investigated detector were tolerated during a measurement with a fixed injector gain. They where corrected for by tuning (by a few volts) the applied HV values after each injector gain setting. Shifts in the injector gain, of the order of 10\%, were considered as a systematic uncertainty of the number of PEs (injected primary electrons). The study of the exact origin of these gain shifts (often considered as being due to charging up effects of the multiplier's substrate~\cite{Breskin10,Tessarotto10})~is beyond the scope of this article.

\section{The methods validation}
\label{sec:Validation}

The charge-injector and the discharge-probability estimation methods described above should be applied with care; e.g. the detector's response should not be affected by the injector's settings, the rate of background discharges should be low compared to that of source-induced discharges, etc. We have performed several experiments to validate the two methods. The experiments, as well as the conditions under which the method can be effectively applied, are discuss in this section.

\subsection{The charge-injector method}
\label{sec:validationInjector}

\paragraph{The transfer gap:}

In the suggested characterization methods both elements, the injector and investigated detector, should remain decoupled - not affecting each other's properties and response. Since they are close to each other, the injector's field may affect that of the investigated detector (and vice-versa), resulting in unwanted gain modifications. This potentially mutual affect depends also on the transfer field between the investigated detector and the injector. It is expected to be small as long as the the investigated detector and the injector are sufficiently distant and the transfer field is no too strong.

The transfer gap between the injector and the detector was set at 5 mm and the transfer field (\Etransfer) was set to 0.5 \kvcm~based on the set of simulations and experiments described below that showed that at this gap the investigated-detector's response is not affected by the injector. Conservatively, the transfer field in these tests was set to 1 \kvcm, enhancing possible mutual affects. 

\begin{figure}
        \centering
        \begin{subfigure}[raggedright]{0.45\textwidth}
               \includegraphics[scale=0.4]{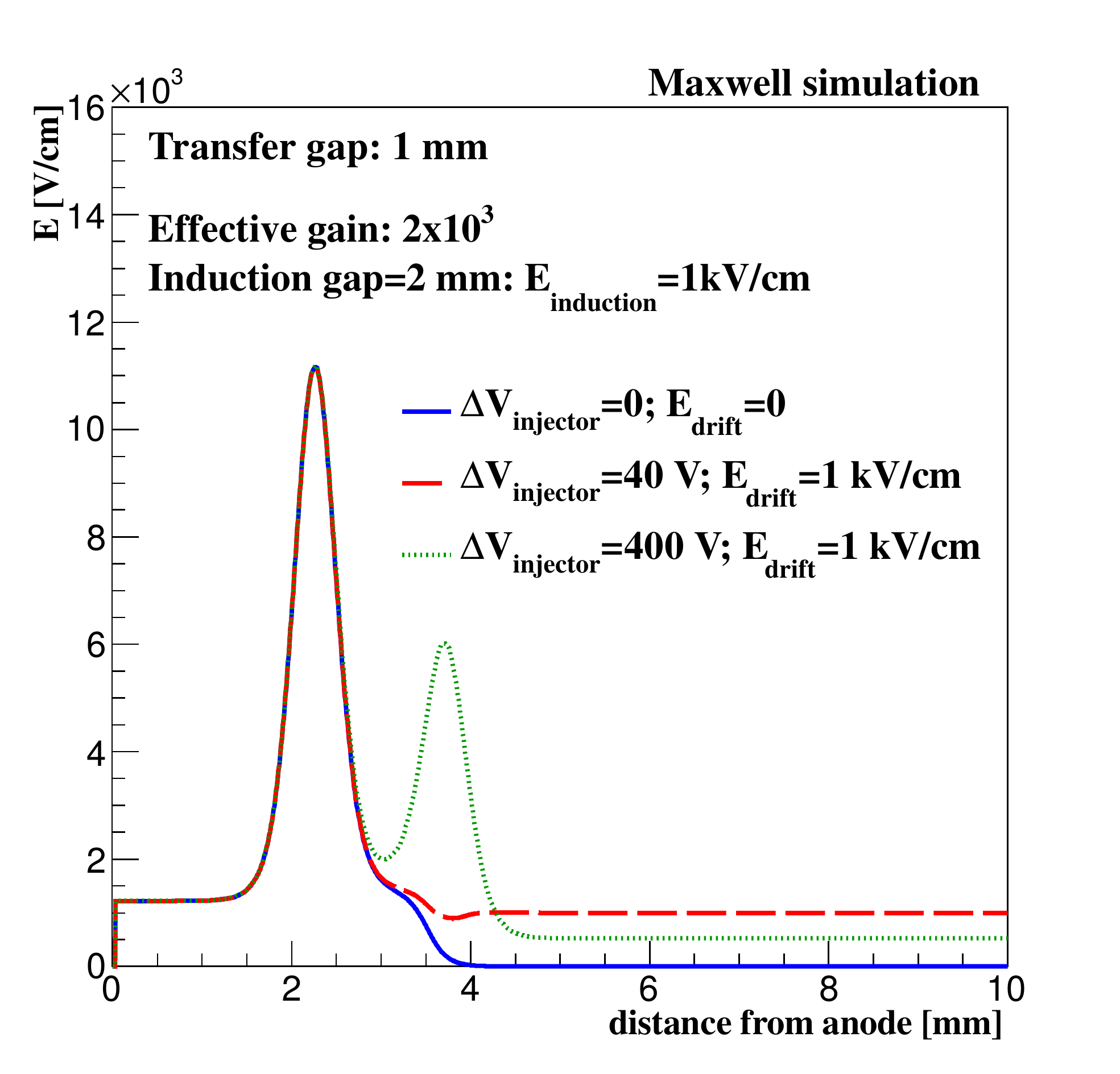}
               \caption{Center of the hole: 1 mm gap}
                \label{fig:1mm}
        \end{subfigure}%
        \quad\quad\quad
        \begin{subfigure}[raggedright]{0.45\textwidth}
                \includegraphics[scale=0.4]{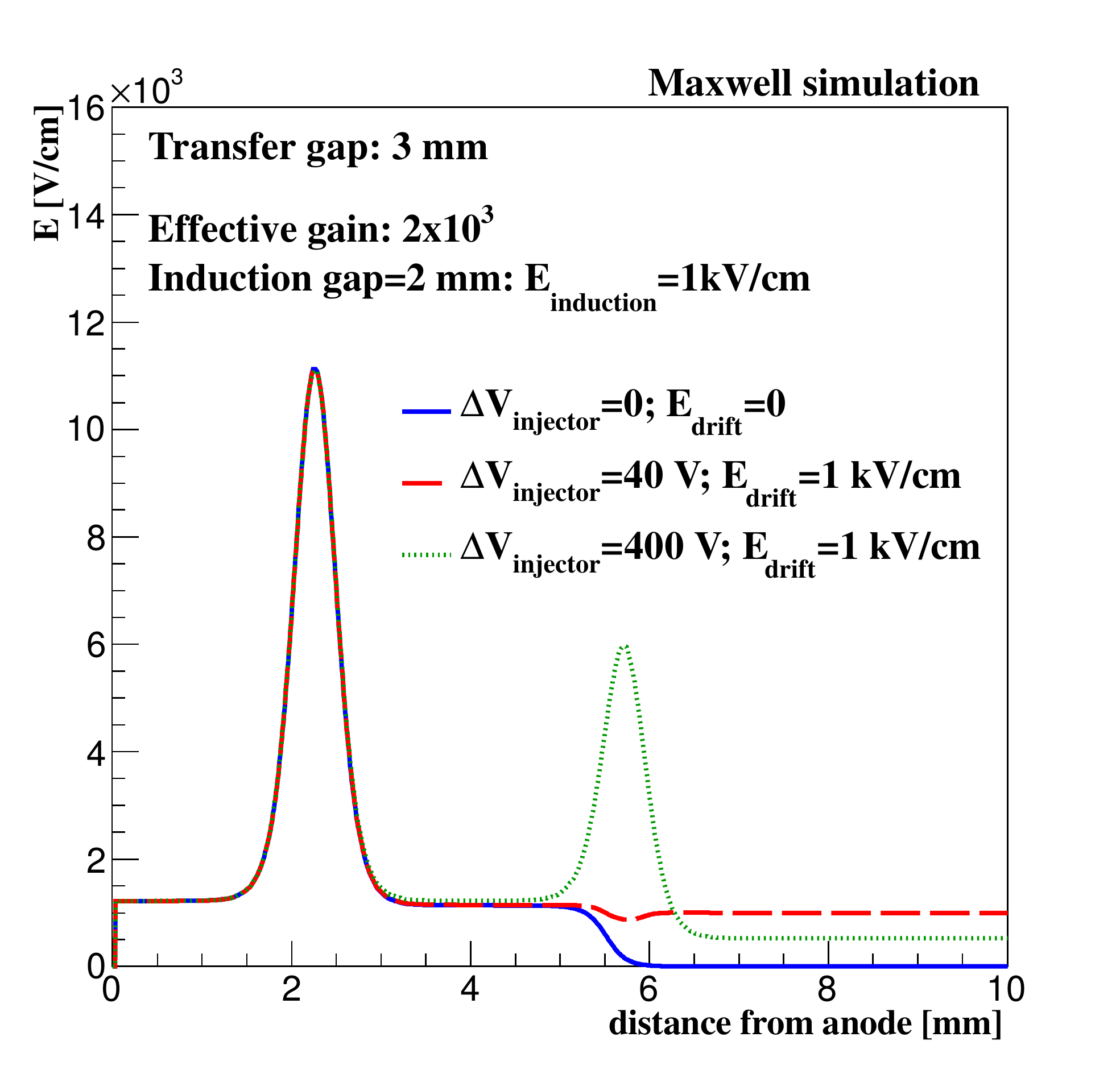}
                \caption{Center of the hole: 3 mm gap}
                \label{fig:3mm}
        \end{subfigure}
		\quad
	    \begin{subfigure}[raggedright]{0.45\textwidth}
                \includegraphics[scale=0.4]{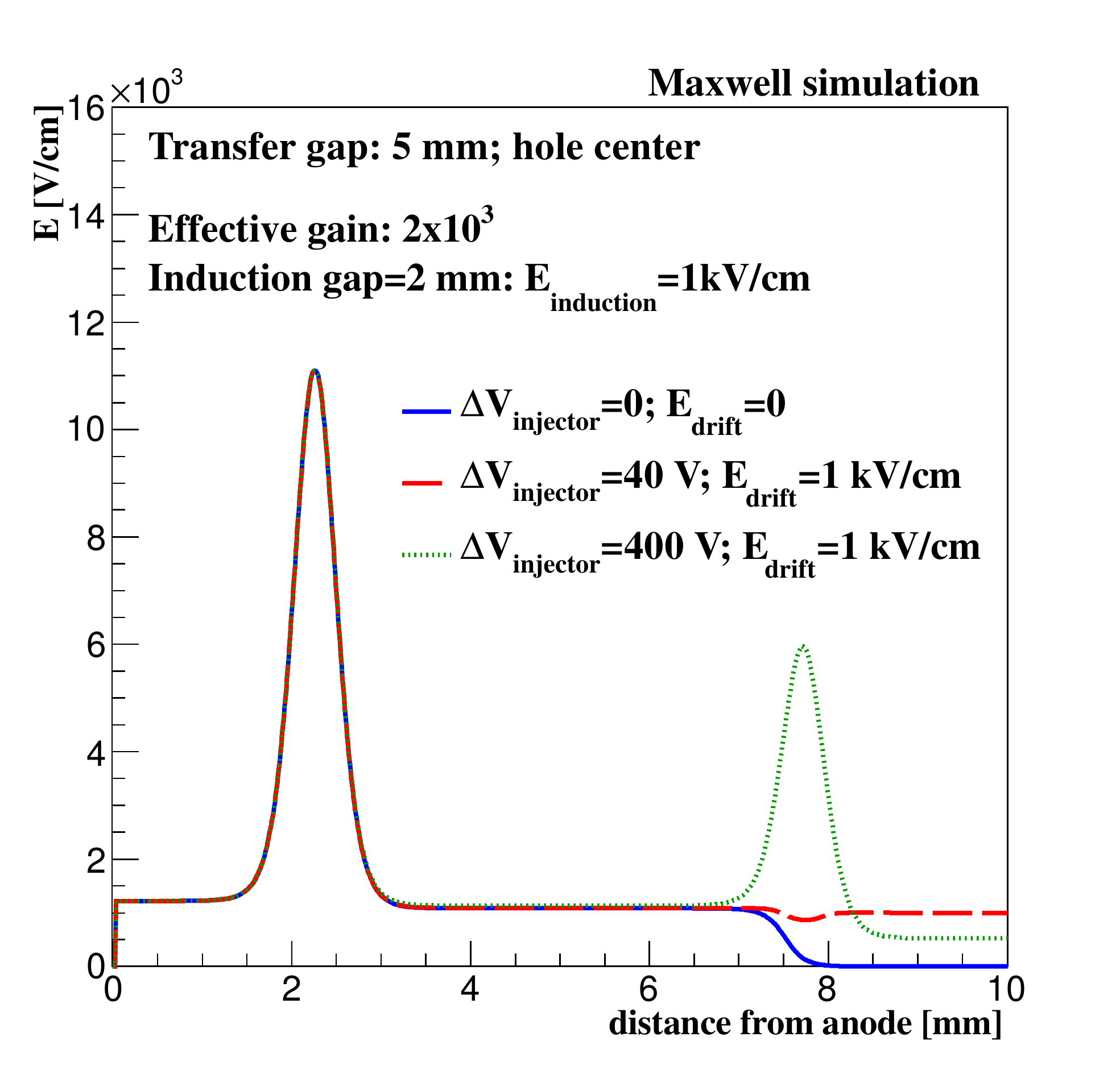}
                \caption{Center of the hole: 5 mm gap}
                \label{fig:5mm}
        \end{subfigure}
        \quad\quad\quad
        \begin{subfigure}[raggedright]{0.45\textwidth}
                \includegraphics[scale=0.4]{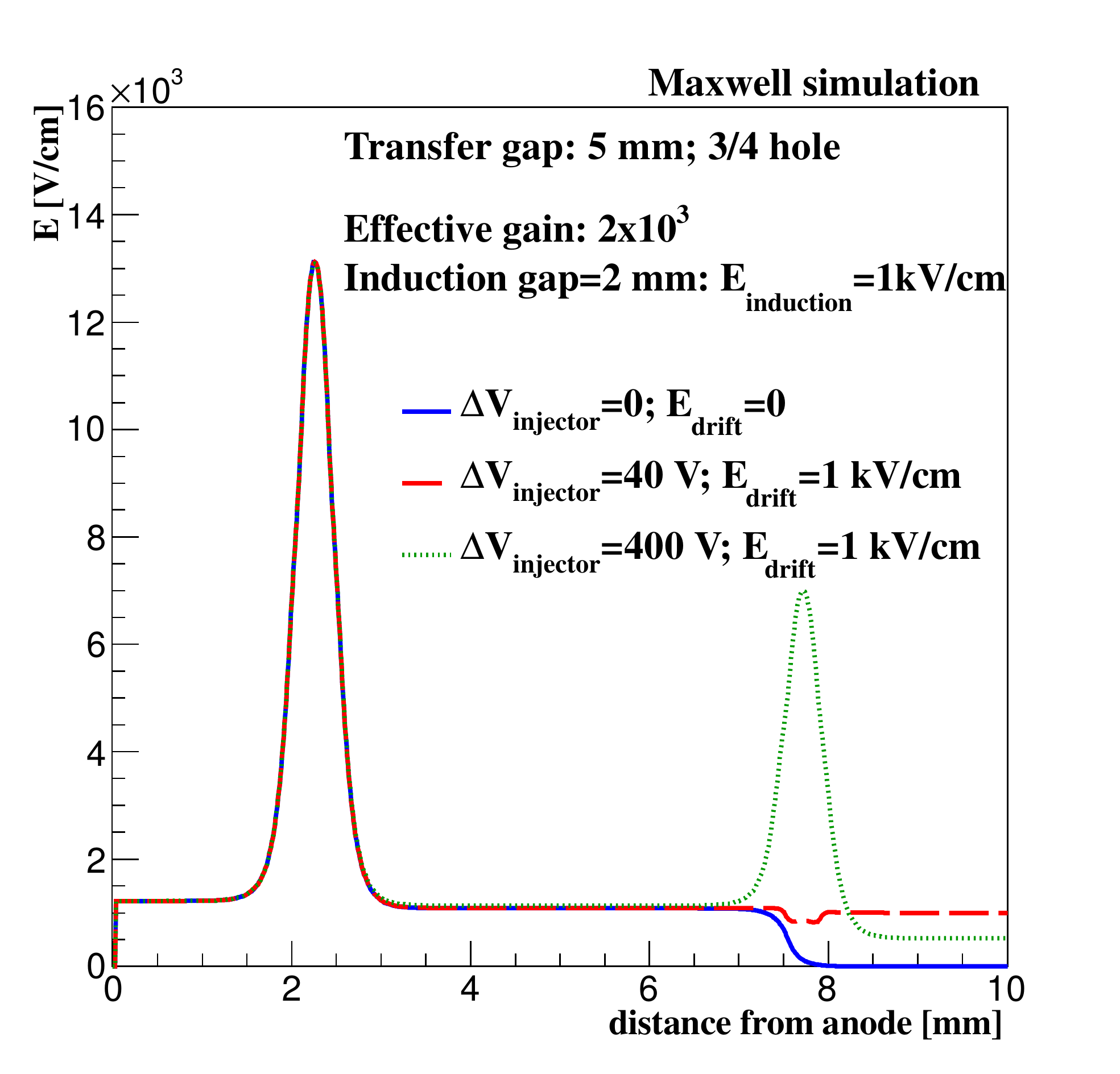}
                \caption{$\mathrm{\frac{3}{4}}$~of hole radius: 5 mm gap}
                \label{fig:5mm34}
        \end{subfigure}
        \caption{Simulation results of the field strength in the direction parallel to the hole-axis. The three curves shown in each figure correspond to different operation conditions of the injector: Solid lines - injector unbiased, \Edrift~= 0, \Etransfer~= 1 \kvcm; dashed lines - $\mathrm{\Delta V_{injector} = 40~V}$, \Edrift~= 1 \kvcm, \Etransfer~= 1 \kvcm ; dotted lines - $\mathrm{\Delta V_{injector} = 400~V}$~operating in multiplication mode (gain $\sim$4), \Edrift~= 1 \kvcm, \Etransfer~= 1 \kvcm. The potential across the investigated THGEM detector was set to 700 V in all cases, corresponding to a gain of $\mathrm{\sim 2 \times 10^3}$, suitable for the detection of MIPs.}
        \label{fig:couplingGarfield}
\end{figure}

Simulation results of the field strength in the direction parallel to the hole axis are shown in figure~\ref{fig:couplingGarfield} for different transfer-gap lengths (positions are measured with respect to the readout anode). As an example we used a configuration of a 0.4 mm thick double-sided THGEM operated with a 2 mm induction gap (\Einduction~= \Edrift~= 1 \kvcm), however the same behavior was recorded with all the configurations investigated here. The detector's and the injector's holes were aligned.

The fields along the hole's axis are shown in figure~\ref{fig:couplingGarfield}a-c for 1, 3 and 5 mm transfer gaps, as function of the distance from anode. The field along a line situated at 3/4 R (R is the hole radius) from the hole axis is shown for the configuration with 5 mm transfer gap (figure~\ref{fig:couplingGarfield}d). The three curves shown in the figures correspond to different operation conditions of the injector: solid-line - $\mathrm{\Delta V_{injector} = 0}$, $\mathrm{\Delta V_{drift} = 0}$, \Etransfer~= 1 \kvcm; dashed line - $\mathrm{\Delta V_{injector} = 40~V}$, \Edrift~= 1 \kvcm, \Etransfer~= 1 \kvcm; dotted line - $\mathrm{\Delta V_{injector} = 400~V}$~(multiplication mode; gain  $\sim4$), \Edrift~= 1 \kvcm, \Etransfer~= 1 \kvcm. The potential across the investigated THGEM detector was set to $\mathrm{\Delta V_{THGEM} = 700~V}$~in all cases, corresponding to a gain of $\mathrm{2 \times 10^3}$, suitable for the detection of minimum ionizing particles~\cite{Arazi12,Bressler13}.

The peak seen right above 2 mm corresponds to the strong dipole field within the THGEM holes. At transfer gaps of 3 and 5 mm, the shape of this peak remains the same regardless of the injector's operation conditions. Indeed, although the fields of the detector and the injector escape the holes, at sufficiently large transfer gaps they do not overlap. At a transfer gap of 1 mm, the fields clearly overlap, indicating that the injector and the detector are coupled. Conservatively, we placed the injector 5 mm away from the investigated detector.

\begin{figure}
        
        \begin{subfigure}[raggedright]{0.45\textwidth}
               \includegraphics[scale=0.4]{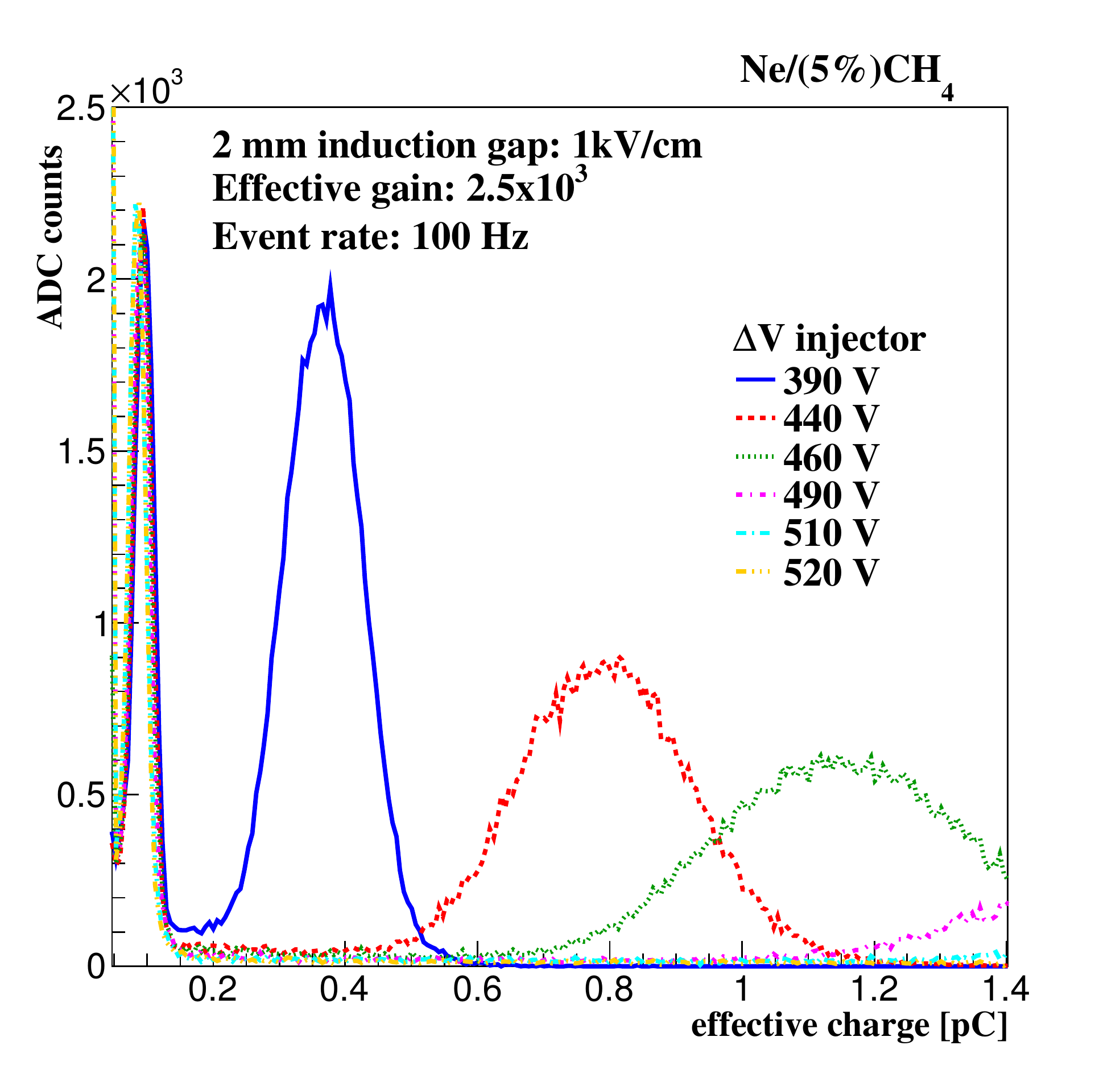}
                \hspace{.1\linewidth}\parbox[l]{.1\linewidth}\caption{}
                \label{fig:spectraDiffInjector}
        \end{subfigure}%
        \quad\quad\quad
        \begin{subfigure}[raggedright]{0.45\textwidth}
                \includegraphics[scale=0.4]{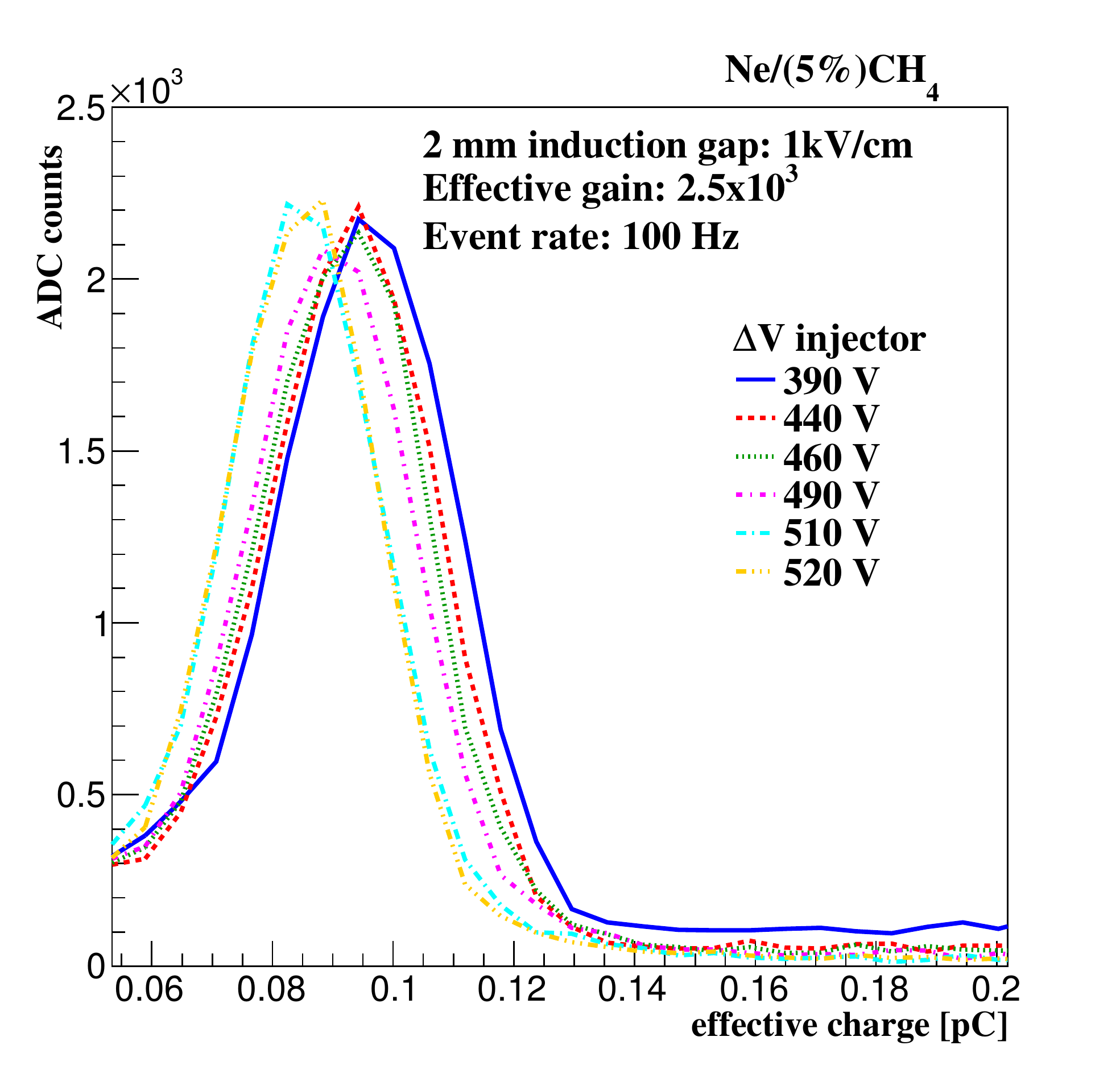}
                 \hspace{.1\linewidth}\parbox[l]{.1\linewidth}\caption{}
                \label{fig:spectraDiffInjectorZoom}
        \end{subfigure}
 \caption{The measured X-ray spectra of a 0.4 mm THGEM detector with 2 mm induction gap preceded by an injector operated with different applied potentials. (a) The full charge-range. (b) Zoom into the charge-range corresponding to events in which the photo-electrons underwent multiplication only in the detector.}
  \label{fig:couplingData}
\end{figure}

The spectra measured at different injector gains are shown in figure~\ref{fig:couplingData}a. The high-charge peak, corresponding to events in which the electrons underwent multiplication in the injector and the detector, is seen only for relatively low injector gains due to the limited range of the MCA. Figure~\ref{fig:couplingData}b shows the same spectra but focuses on the low-charge peak that corresponds to events in which the electrons underwent multiplication only in the investigated detector. A mild (less than 10\% in total) monotonic decrease of the gain was observed when increasing the injector gain. We conclude that when the injector is placed 5 mm away from the investigated detector, its effect on the detector response (e.g. on its gain) is very small.

\paragraph{The injector gain range:}

Our investigated detectors were operated at a fixed gain (\effgain) and with constant drift and transfer fields (\Etransfer~= \Edrift~= 0.5 \kvcm). Under these conditions, the rate of background discharges may have varied with the injector gain. In particular, at high injector gains discharge could occur in the injector itself. A meaningful measurement of the discharge probability requires an accurate estimation of the rate of the background discharges. Therefore, it is important to work in a range of injector gains over which the background-discharges rate is stable and small. 

\begin{figure}[h]
  \centering
  \includegraphics[scale=0.5]{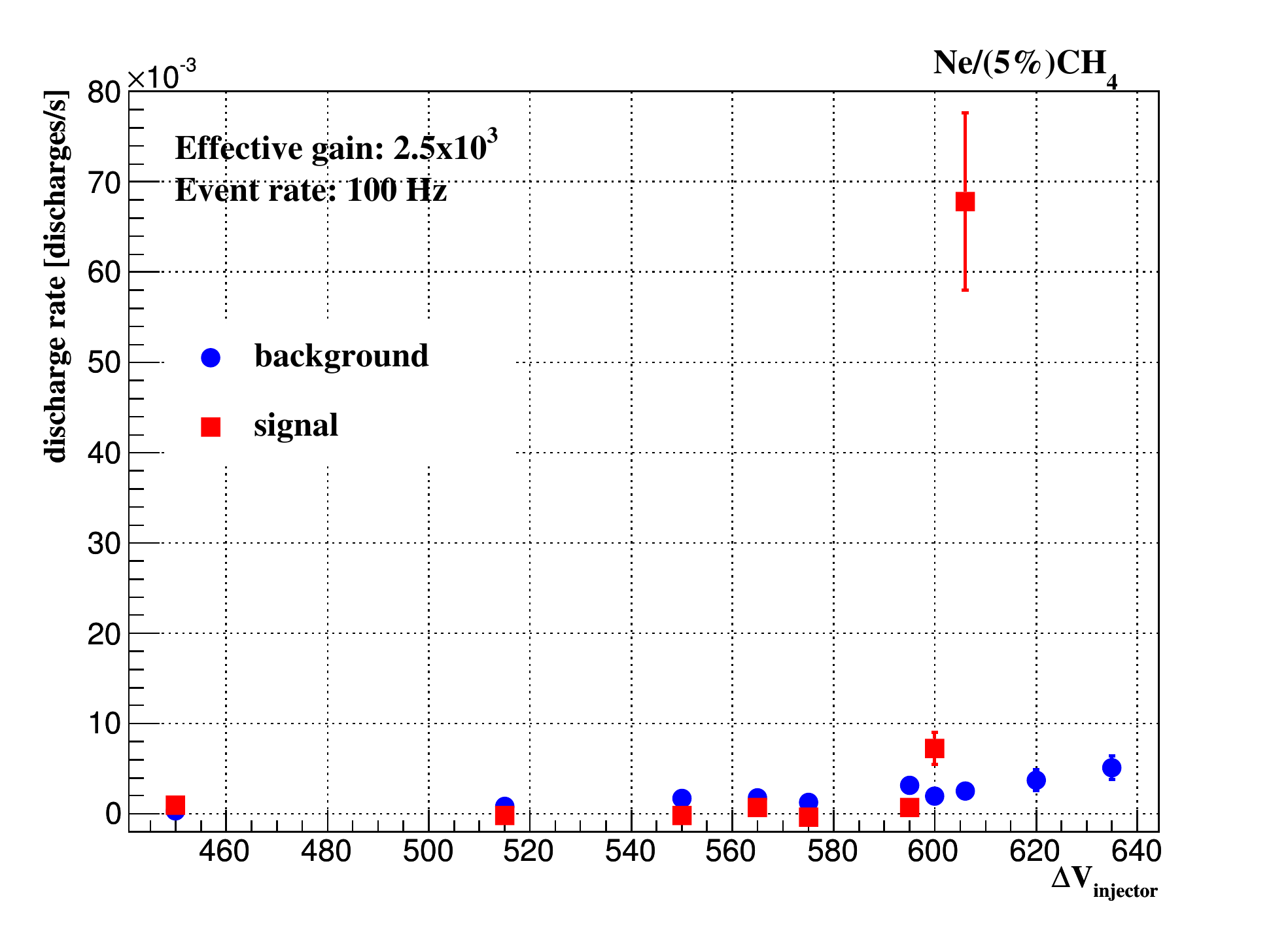}
  \caption{The rate of the background discharges as a function of the injector voltage, and that of discharge measured with a source in a 0.4 mm double-sided THGEM with 1 mm induction gap without gap multiplication (\Etransfer~= \Edrift~= 0.5 \kvcm, \Einduction = 1 \kvcm).}
  \label{fig:maxInjectorGain}
\end{figure}

Figure~\ref{fig:maxInjectorGain} shows the rate of the background discharges as a function of the injector voltage (blue round markers), and the rate of discharge measured in the same configuration (red square markers). The measurements where performed with a 0.4 mm double-sided THGEM with 1 mm induction gap (\Einduction~= 1 \kvcm). As can be seen, over the injector gain range of $\sim$4 and $\sim$37 (PEs range between $\sim$750 and $\sim$7000, $\mathrm{\Delta V_{injector} = 450~V}$~and 600 V respectively) the background discharge rate is almost constant. A small increase is observed at gains above $\sim$37 ($\mathrm{\Delta V_{injector} \geq 600~V}$). The injector-gain range in this configuration is, therefore, limited to values below that value. This range of the injector gain was sufficient to evaluate the dynamic-range of this configuration since when it was operated with constant source irradiation a rapid increase was observed in the discharges rate, at an injector gain of $\sim$34 ($\sim$6400 PEs, $\mathrm{\Delta V_{injector} = 570~V}$), below the limitation imposed by the rate of background discharges.

A similar behaviour, namely a sharp increase in the discharge probability and a stable background discharge rate, was observed in all the configurations investigated. The increase in the discharge probability could indicate upon the avalanche exceeding of the Raether limits. Since the PE distribution does not have a high-charge tail (figure~\ref{fig:bremsstrahlung}), the transition from stable operation to a discharge-regime occurs over a narrow range.

\subsection{Discharge probability measurement}

\paragraph{Counting rate dependence:}

The gain of gaseous detectors, including that of THGEMs~\cite{Peskov10,Arazi13_2}, varies with the local counting rate; other detector properties, such as the discharge probability should also be rate-dependent. In this study, we wanted to avoid uncertainties due to possible dependence of the discharge probability on the counting rate. 

\begin{figure}[h]
  \centering
  \includegraphics[scale=0.5]{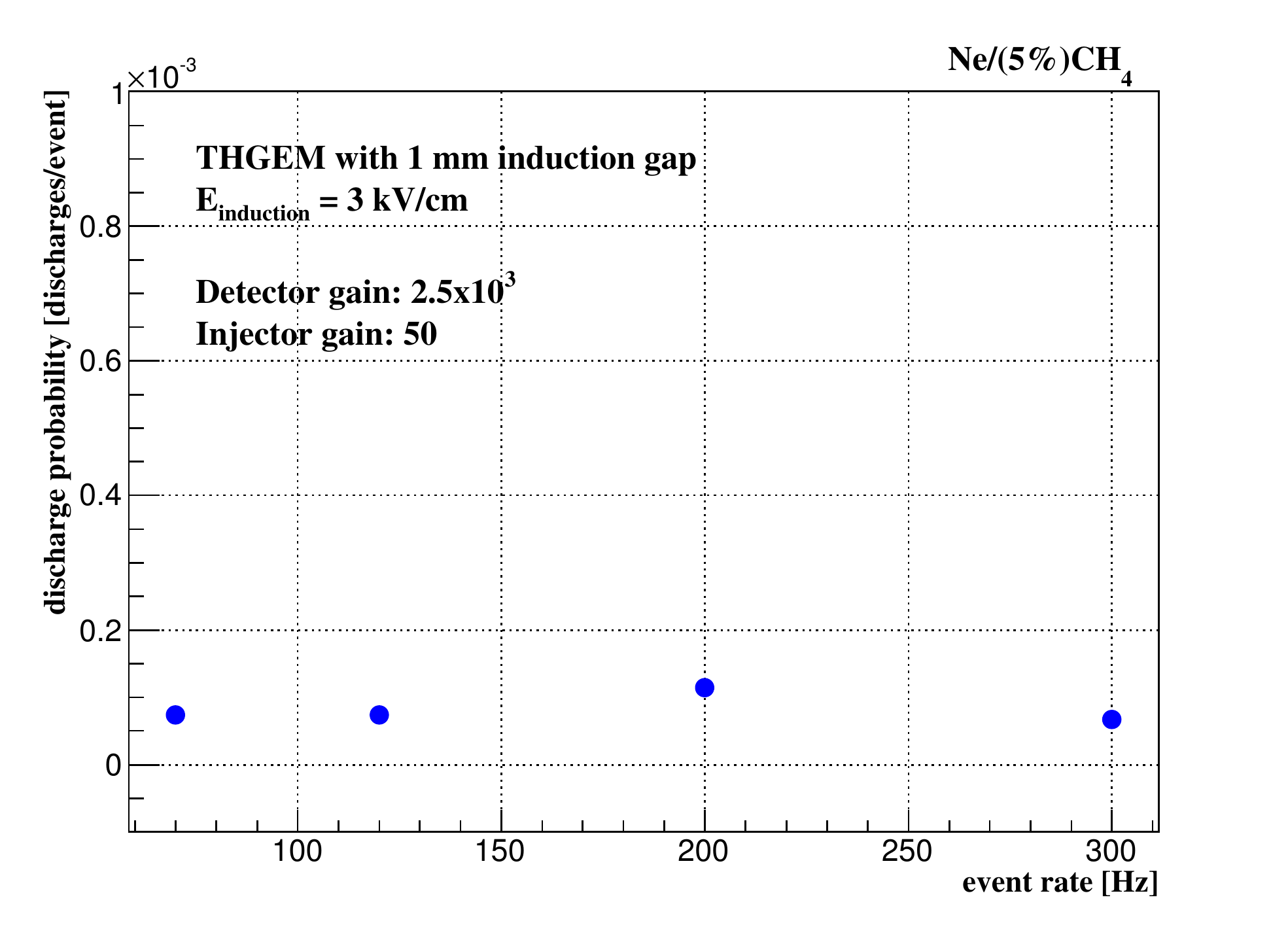}
  \caption{The discharge probability a 0.4 mm THGEM with 1 mm induction gap (\Einduction~= 3 \kvcm; \Edrift~= \Etransfer~= 0.5 \kvcm) as a function of the irradiation rate. The detector was operated at an effective gain of \effgain~and the injector gain was fixed at 50.}
  \label{fig:dischargesVsRate}
\end{figure}

The discharge probability as a function of the counting rate is shown in figure~\ref{fig:dischargesVsRate} over the range of 70 to 300 Hz. For illustration, we used a 0.4 mm THGEM with 1 mm induction gap operated with multiplication in the induction gap (\Einduction~= 3 \kvcm; \Edrift~= \Etransfer~= 0.5 \kvcm). Note that a similar behaviour was recorded with all the configurations investigated in this work. This investigated THGEM detector was operated at an effective gain of \effgain; the injector gain was set at 50 ($\sim 9.4 \times 10^3$~PEs).

As can be seen in figure~\ref{fig:dischargesVsRate}, the difference in the discharge probability over the investigated counting rate is within the uncertainty of the measurement. The measurements in this work were conducted at rates of the order of 100 Hz; they were not affected by eventual small rate fluctuations.

\paragraph{The Poisson-distribution of the discharges:}

\begin{figure}[h]
  \centering
  \includegraphics[scale=0.5]{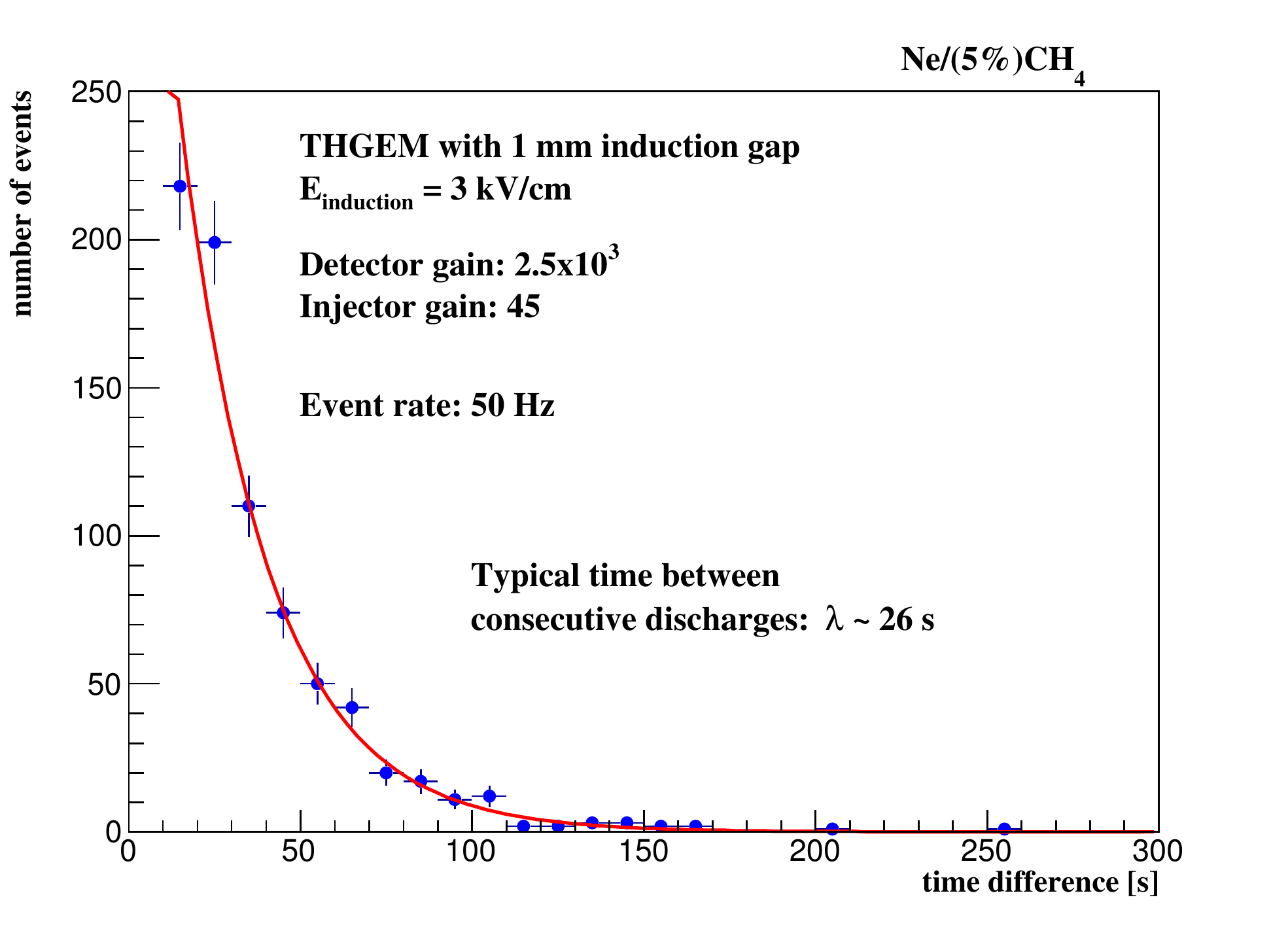}
  \caption{The time difference between consecutive background discharges; the solid line is the exponential fit to the histogram.}
  \label{fig:Poisson}
\end{figure}

When calculating the discharge probability we assumed that consecutive background and signal discharges are not correlated in time. If, for instance, the origin of the discharge was up-charging of a specific point in the detector, with a characteristic time T seconds, a discharge would have occurred every $\sim$T seconds. Hence, we assumed that the number of discharges follows a Poisson distribution. We confirmed this assumption exploiting the Poisson-exponential relation and showed that the time between each pair of consecutive discharges has, indeed, an exponential distribution. The results are demonstrated in figure~\ref{fig:Poisson}: the histogram (blue dots) summarizes the measured time difference between consecutive discharges. The solid line is the exponential fit to the histogram.

\paragraph{Discharge probability of similar electrodes:}

In an attempt to begin standardizing the process of electrode characterization we selected electrodes with low leak-current values ($\mathrm{\leq 5 \times 10^{-11}~A}$~at 1 kV in air) and with breakdown voltages not lower than those predicted by the Paschen law~\cite{Paschen1889}. Uncertainties arising from the production itself were avoided by using electrode made by a single producer using the same production technique.

\begin{figure}[h]
  \centering
  \includegraphics[scale=0.5]{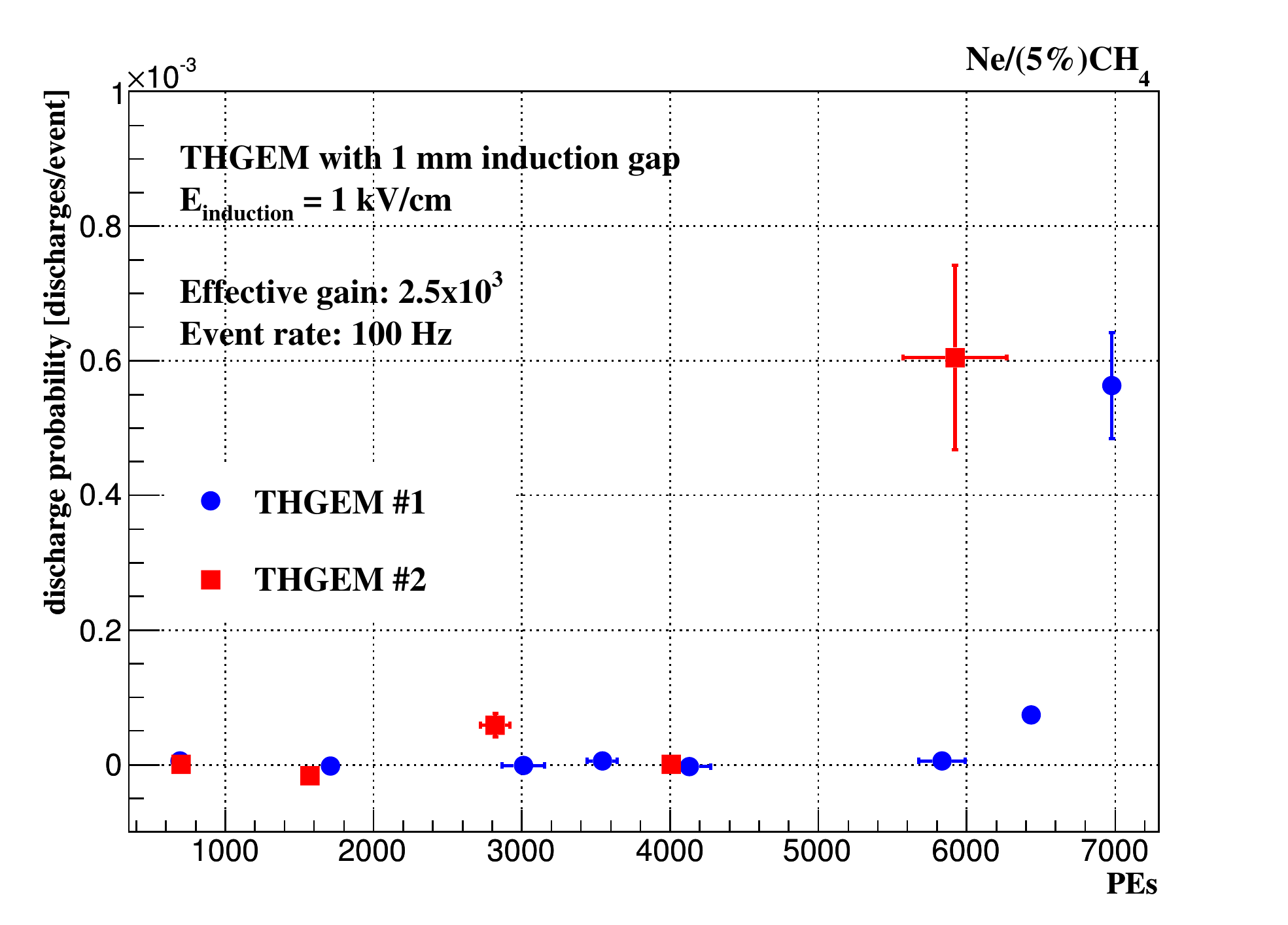}
  \caption{The discharge probability of two similar 0.4 mm thick double-sided THGEM detectors, with 0.5 mm hole diameter, 1 mm pitch and 100 $\mathrm{\mu m}$ , operated with a 1 mm induction gap (\Einduction~= 1 \kvcm, \Etransfer~= \Edrift~= 0.5 \kvcm). The electrical properties of the electrodes are described in the text.}
  \label{fig:electrodes}
\end{figure}

We compared the discharge probability of electrodes with identical geometry (0.4 mm thick double-sided THGEM, with 0.5 mm hole diameter drilled in an hexagonal lattice with a pitch of 1 mm and 0.1 mm rim) operated in the same configuration (1 mm induction gap; \Einduction~= 1 \kvcm). Their measured leakage-current values under 1 kV in air were both $\mathrm{5 \times 10^{-11}~A}$. The maximum sustainable voltage in air (defined here as the voltage in which the time difference between consecutive discharges is less than 10 seconds) of the first electrode was 3780 V and that of the second was 3800 V. As can be seen in figure~\ref{fig:electrodes}, the response of the two electrodes is rather similar. In particular, the sharp increase in the measured discharge probability occurred at a similar primary charge value of $\mathrm{5 \times 10^3}$~PEs. Similar behaviours of the discharge probabilities were also observed with electrodes of equal parameters operated in the other configurations.

This result, though recorded only with two electrodes, suggests that the discharge-probability measured with a given electrode is applicable to others, of the same structure and of similar electrical properties, if made by the same producers.

\section{Results: Instabilities evaluation in THGEM-structures}
\label{sec:Results}

We present the results of a comparative study, carried out with the methods described above, of the discharge probability of several THGEM-structures: double-sided THGEM, Thick WELL (THWELL~\cite{Arazi12}) and Resistive Thick WELL (RWELL~\cite{Arazi12,Arazi13_2}). We used electrodes with 0.5 mm hole-diameter and 1 mm pitch. All the measurements were conducted at a total detector effective gain of \effgain. The drift and transfer fields were kept constant (\Edrift~= \Etransfer~= 0.5 \kvcm). The configurations are summarized in table~\ref{tab:configurations}.

\begin{table}[h]
 \caption{The different THGEM-structures.}
   \centering
    \begin{tabular}{ l l c  c c c  r }
    \hline
     & 	Structure  &	Thickness & rim & Induction gap  & Induction field	 &	Resistivity                     \\ 
    		   & & [mm] & [$\mathrm{\mu m}$] &	[mm]		& [\kvcm] &  \textrm{$\mathrm{[\frac{M\Omega}{square}]}$}  \\
    \hline
    \textit{conf1} & THGEM	 & 0.4  & 100 & 1 			&	1		&		-     	\\
    \textit{conf2} &  THGEM	 & 0.4  & 100 & 	1 			&	3		&		-     	\\
    \textit{conf3} &  THGEM	 & 0.4  & 100 &  1 			&	5		&		-     	\\
     \textit{conf4} &  THGEM	 & 0.4  & 50 &  	1 			&	5		&		-     	\\
    \textit{conf5} &     THWELL	 & 0.4 & 100 & - 			&	-		& 	      -     	\\
    \textit{conf6} &     RWELL   	& 0.4 & 100 &	-			&	-		&		1   		 \\
    \textit{conf7} &     RWELL   	& 0.4  & 100 &  	-			&	-		&	      10   	 	\\ 
    \textit{conf8} &     RWELL   	& 0.8  & 100 & 	-			&	-		&	      10   	 	\\    
    \label{tab:configurations}
    \end{tabular}
\end{table}  

\subsection{Double-sided THGEM with an induction gap}

\begin{figure}[h]
  \centering
  \includegraphics[scale=0.5]{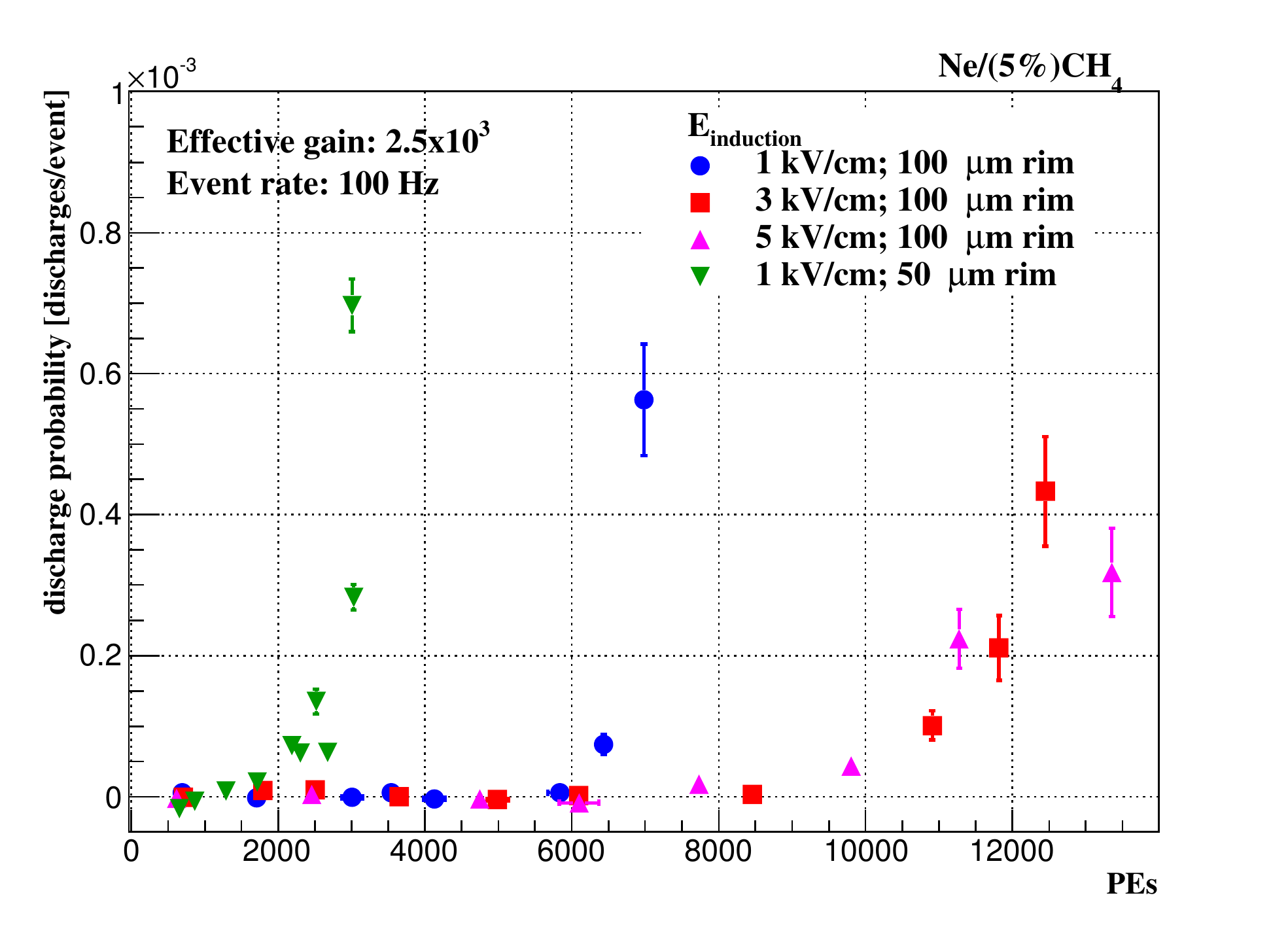}
  \caption{The discharge probability of a 0.4 mm double-sided THGEM with 1 mm induction gap as a function of the number of primary electrons. The configuration (table~\protect\ref{tab:configurations}), the rim size (h) and the \Einduction~value for each graph are given in the figure.}
  \label{fig:resultsTHGEMinduction}
\end{figure}

We studied the dynamic-range of 0.4 mm thick standard double-sided THGEM structures operated with 1 mm induction gap. The investigations aimed at comparing configurations with hole-multiplication only to that with charge multiplication in both the THGEM holes and the induction gap;  the rim-size effect on the discharge probability was studied as well. The discharge probability of several detector configurations is compared in figure~\ref{fig:resultsTHGEMinduction}. The configurations and the conclusions are discussed below.

In \nech, the onset of electron multiplication is at E $\sim$2.0 \kvcm~\cite{Azevedo2010}. Figure~\ref{fig:resultsTHGEMinduction} depicts discharge probability results investigated with induction-field values of \Einduction~= 1 \kvcm~(no multiplication in the induction-gap) and \Einduction~= 3 and 5 \kvcm; the respective gain values within the induction gap under these fields are $\sim$10 and $\sim$100. These gains were estimated from the ratio between the total gain (\effgain, measured from the MCA spectra) and the gain within the THGEM holes. The low-gain values within the THGEM holes ($\sim$250 and $\sim$25 respectively) were measured in the way described above for the low injector gains. 

As can be seen, the dynamic-range of the configurations sharing the total multiplication between the detector and the induction gap is broader; the onset of discharge-probability rise with gap multiplication is at $\mathrm{\sim 9 \times 10^3}$~primary electrons, compared to $\mathrm{\sim 5 \times 10^3}$ primary electrons with THGEM multiplication only. This result is consistent with our expectations; the final steps of multiplication occur within the induction gap where the field is lower than inside the holes and the geometry is broader, resulting in a higher Reather limit~\cite{Raether} and larger dynamic-range.

The dynamic range of a THGEM with a hole-rim of 50 $\mathrm{\mu m}$~(\Einduction~= 1 \kvcm) is also shown in figure~\ref{fig:resultsTHGEMinduction}. The dynamic-range of this configuration was narrower than that with a 100 $\mathrm{\mu m}$~rim, with the discharge probability rising already at $\mathrm{\sim 10^3}$~primary electrons. The result is in agreement with the observation that higher gains are reachable with THGEM electrodes of larger hole-rims~\cite{Alexeev09}.

\subsection{WELL-structures}

\begin{figure}[h]
  \centering
  \includegraphics[scale=0.5]{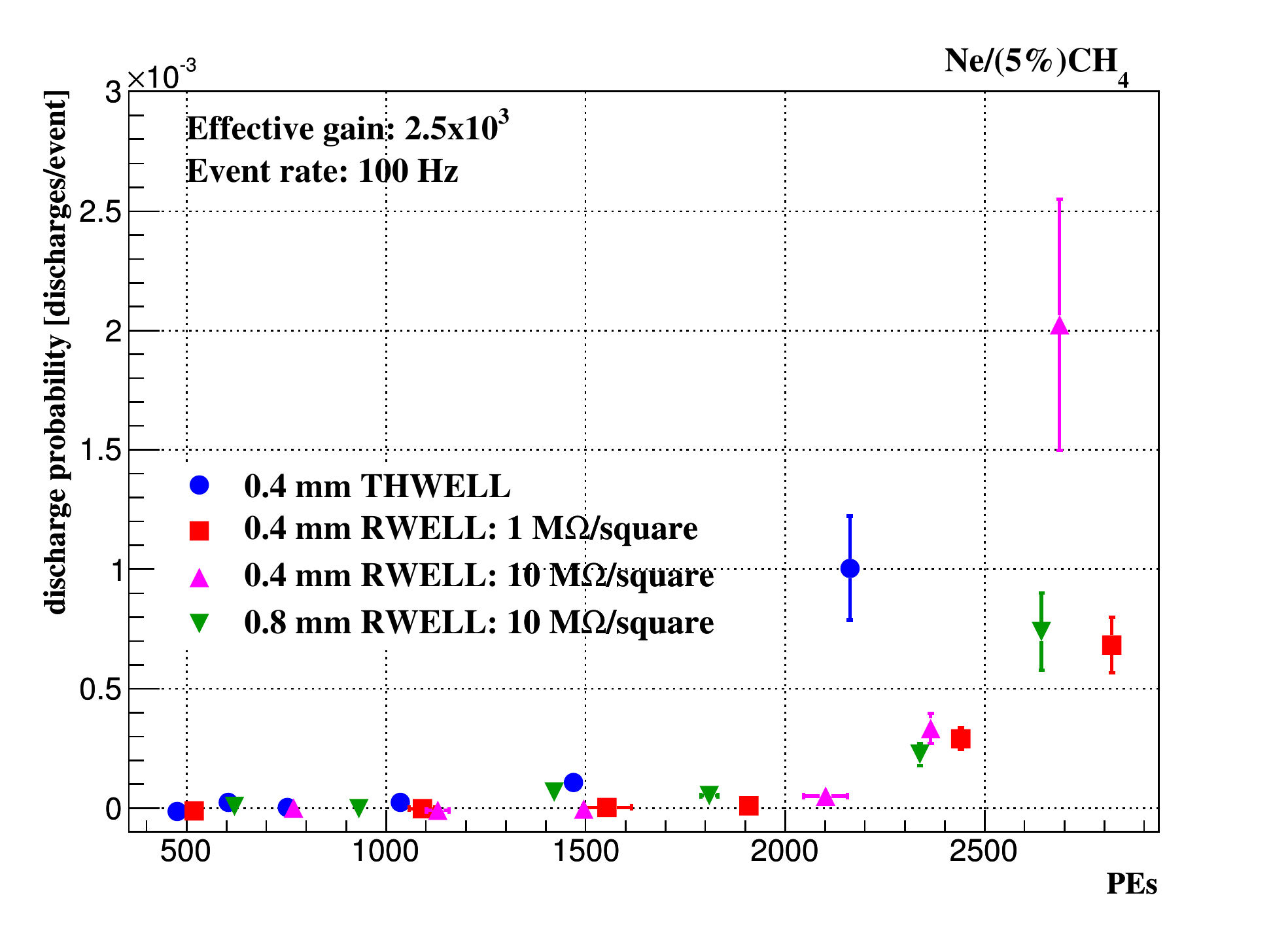} 
  \caption{The discharge probability as a function of the number of primary electrons of 0.4 mm THWELL (\textit{conf5}), 0.4 mm RWELL with surface resistivity of 1 \mos~(\textit{conf6}),  0.4 mm RWELL with surface resistivity of 10 \mos~(\textit{conf7}) and  0.8 mm RWELL with surface resistivity of 10 \mos~ (\textit{conf8}). The configurations are described in table~\protect\ref{tab:configurations}. }
  \label{fig:resultsWELL}
\end{figure}

We studied the effect of the resistive layer and the electrode thickness on the dynamic-range of
WELL-structures. The discharge probability of 0.4 mm thick THWELL and RWELL (with surface resistivity values of 1 and 10  \mos) and that of a 0.8 mm thick RWELL (with surface resistivity of 10  \mos) are compared in figure~\ref{fig:resultsWELL}.

The dynamic range of the different RWELL configurations is slightly broader than that of the THWELL. It was shown in~\cite{Arazi13_2}, that the typical discharge energy is quenched in RWELL configurations compared to that in THWELL ones. However, the possible role of the resistive film in decreasing somewhat the discharge probability is yet unclear and is the subject of further studies.

A similar dynamic range is observed in figure~\ref{fig:resultsWELL} in both 0.4 and 0.8 mm thick RWELL detectors (with 10 \mos~anodes). As the last multiplication steps occur close to the bottom anode, and as both RWELL detectors were operated at equal effective gains, with a weaker field in the thicker electrode - we would have expected a higher Raether limit for the latter and consequently a broader dynamic-range.

\section{Conclusions and discussion}
\label{sec:Conclusion}

A method was systematically investigated to mimic highly ionizing particles (HIPs) in the laboratory in a controlled way; its usefulness for gas-avalanche detector studies was demonstrated. In particular, unlike detector studies with HIPs in hadronic and other accelerator beams or sources, we showed the possibility of producing rather narrow X-ray induced tuned primary-electron distributions with precise mean value and no high-charge tails. We demonstrated the validity of the method on THGEM detectors - though it is applicable to a broad range of other gas-avalanche radiation detectors.
A methodology for measuring discharge probabilities was described in detail. Attention was given to external effects, such as unrelated background discharges and detector and power-supply dead times.

The studies revealed the existence of background discharges occurring in the absence of a radiation source - at rates higher than expected from cosmic radiation. The background discharges were observed in all the configurations investigated, regardless of the electrode thickness, geometry and applied injector voltage. Understanding their origin, which could be related to the natural radioactivity of FR4, is the subject of other ongoing studies.

The response of different THGEM-structures, double-sided THGEM and single-sided WELL (a THGEM with closed bottom) to events with a broad spectrum of primary ionization was investigated. The studies concentrated on thin detector configurations operated at gains of order \effgain, suitable for the detection of minimum ionizing particles. 


A typical drift gap of a thin THGEM-structure is of order 3-5 mm. In \nech~(were 150 GeV MIPs deposit on the average 60 electrons/cm) a MIP is expected to deposit 20-30 PEs. In this work we have studied the response of the investigated detectors to energy deposits 10 to 400 times larger than the typical energy deposited by MIPs. We were able to associate the discharges with a specific energy deposition to a good level of accuracy.

We found out that double-sided THGEM-structures could tolerate HIPs which deposit 300 times more energy compared to MIPs, when the effective gain is divided between multiplication in the THGEM and within the induction gap preceding the anode. We also found that the dynamic-range of a THGEM with a gain of 10 to 100 in the induction gap was about 4-fold larger than that of a RWELL with 10 \mos~dresistive anode - both operating at an equal gain of \effgain. Indeed, in the double-sided configuration part of the avalanche develops within the induction gap, where the field is weaker and the avalanche density is smaller; this could lead to a higher effective Raether limits, and hence to a broader dynamic-range. It was confirmed again that THGEM electrodes with larger rims are more stable then those with smaller ones. Resistive anodes were found to slightly broaden the dynamic-range of WELL-structures; the difference in dynamic range between a resistivity of 1 \mos~and 10 \mos~was negligible.

The results obtained in this work provide us with a better understanding of the expected performance of THGEM structures. They demonstrate the usefulness of the charge-injector method that permitted a better comparison between different THGEM configurations.

\begin{figure}[h]
  \centering
  \includegraphics[scale=1]{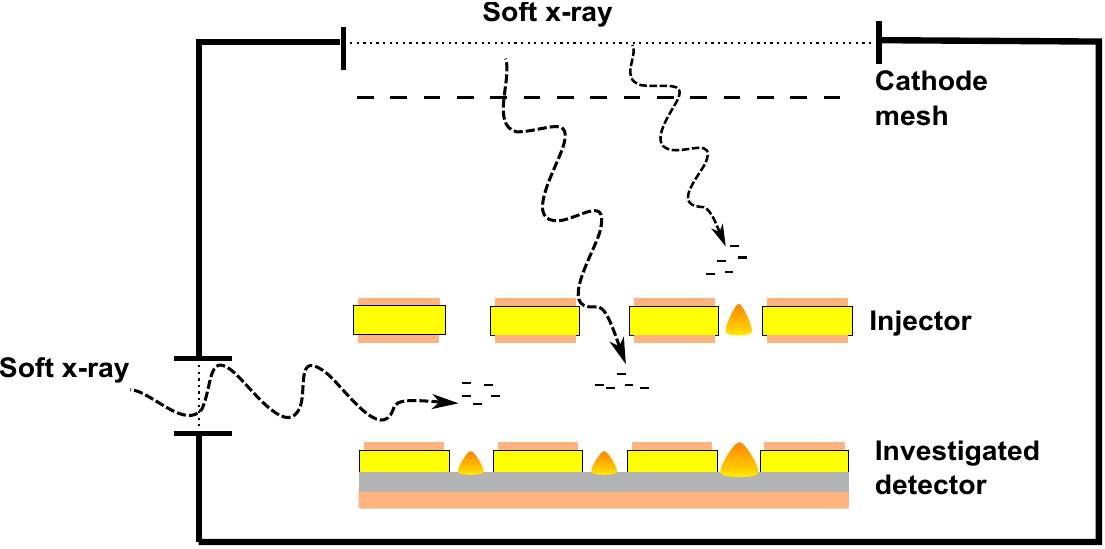}
  \caption{Schematic description of a setup with simultaneous irradiation of a detector with independently rate- and charge-tuned MIP-like and HIP-like sources.}
  \label{fig:mixed_source}
\end{figure}

The charge-injector method could be further elaborated, as depicted in figure~\ref{fig:mixed_source}; it allows for irradiating the detector simultaneously, with two sources. The side source, delivering charges multiplied only in the investigated detector, would mimic MIPs at tunable rates; the on-axis source would yield a tunable number of primary electrons (multiplication in the injector and in the investigated detector) mimicking HIPs. Such setup will make it possible to study, for instance, the response of detectors to high MIPs rates in the presence of rate- and charge-tunable highly ionizing events.

\acknowledgments

This research was supported in part by the I-CORE Program of the Planning and Budgeting Committee and The Israel Science Foundation (grant NO 1937/12) and by the Israel-USA Binational Science Foundation (Grant 2008246). A. Breskin is the W.P. Reuther Professor of Research in the Peaceful use of Atomic Energy.

\appendix
\appendixpage
\addappheadtotoc
\section{Estimating the width of the injector gain} 
\label{appendixA}

The width of the distribution of the injector gain was estimated from a toy Monte-Carlo using the measured distributions \GD(\muD,\sigmaD) and \GT(\muT,\sigmaT). As suggested in~\cite{Rubin_opt13}, we assumed that the charge in each multiplication stage was divided between several holes. We further assumed that in each multiplication stage, the gain in each hole is described by the same Gaussian distribution, and that the mean gain value in each hole is equal to that of the multiplication stage.

We defined the following parameters: \NHinj~and \NHD~- the number of active holes in the injector and the detector respectively. \GHinj(\muHinj,\sigmaHinj) and \GHD(\muHD,\sigmaHD) - the gain distribution in a single injector and detector hole respectively. $\mathrm{G_X}$~stands for a Gaussian distribution with a mean value $\mathrm{\mu_X}$~and a width $\mathrm{\sigma_X}$.

As discussed in section~\ref{sec:Setup}, \GD(\muD,\sigmaD) and \GT(\muT,\sigmaT) are extracted directly from the measured spectra and $\mathrm{\muinj = \frac{\muT}{\muD}}$~is also known. The remaining seven parameters (\sigmainj, \NHinj, \NHD, \muHinj, \sigmaHinj, \muHD~and \sigmaHD) are subject to five constraints:

\begin{enumerate}
\item{\muHinj~= \muinj}
\item{\muHD~= \muD}
\item{\NHinj~and \sigmaHinj~are constrained by \sigmainj: The width of the distribution obtained when summing \NHinj~random numbers drawn from a Gaussian distribution with a width \sigmaHinj~should equal \sigmainj}.
\item{\NHD~and \sigmaHD~are constrained by \sigmaD: The width of the distribution obtained when summing \NHD~random numbers drawn from a Gaussian distribution with a width \sigmaHD~should equal \sigmaD}
\item{\NHinj, \sigmaHinj, \sigmainj, \NHD, \sigmaHD~and \sigmaD~are constrained by \GT(\muT,\sigmaT)}
\end{enumerate}

Since the model has two degrees of freedom (seven parameters and five constraints), it has more than one solution. Advanced statistical analysis selecting the most likely fit parameters, as well as possible techniques to find additional two constraints are beyond the scope of this work. 

As an example, different points in the parameter space of the model that fit well a spectrum of a double-sided THGEM with 1 mm induction gap and an injector operated at a mean gain of 19 are summarized in table~\ref{tab:MC}. As can be seen, \sigmaHinj~increases with \NHinj~and it is independent of the other parameters.  Based on the measured hole multiplicity in a single-stage THGEM detector~\cite{Rubin_opt13}, we conservatively selected \NHinj~= 4, obtained \sigmaHinj~and calculated \sigmainj. In this example (\sigmaHinj~= 22\%) the resulting width of the injector gain is 11\%. Conservatively, we fixed the upper bound on the width of the distribution of the injector gain to be 15\%.

\begin{table}[h]
 \caption{Different points in the parameter space of the Monte-Carlo model that fit well a spectrum of double-sided THGEM with 1 mm induction gap (\Einduction~= 1 \kvcm) and injector operated at a mean gain of 19.}
   \centering
    \begin{tabular}{c c c c}
    \hline
\NHinj & \NHD & \sigmaHD~[\%] & \sigmaHinj~[\%]  \\
\hline
2   & 4   & 31   & 10    \\
3   & 4   & 31   & 18    \\
4   & 4   & 33   & 22    \\
2   & 5   & 37   & 10    \\
3   & 5   & 37   & 18    \\
4   & 5   & 37   & 22    \\    
    \label{tab:MC}
    \end{tabular}
\end{table}


\begin{thebibliography}{9}

\bibitem{Fonte2001}
C. Iacobaeus, M. Danielsson, P. Fonte, T. Francke, J. Ostling, and V. Peskov, \emph{Sporadic electron jets from cathodes - the main breakdown-triggering mechanism in gaseous detectors}, {\emph{ IEEE Nucl. Sci. Symp. Conf. Rec.} {\bf 1 525} (2001)}

\bibitem{Fonte1997}
P. Fonte, V. Peskov, and B. D. Ramsey, \emph{Streamers in MSGC's and other gaseous detectors}, {\emph{ICFA Instrum.Bull.}} {\bf{15 1} (1997)}

\bibitem{Peskov10}
V. Peskov, M. Cortesi, R. Chechik and A. Breskin \emph{Further evaluation of a THGEM UV-photon detector for RICH-comparison with MWPC}, \jinst{5}{2010}{P11004}

\bibitem{Breskin09}
A. Breskin et al., \emph{A concise review on THGEM detectors},\href{http://www.sciencedirect.com/science/article/pii/S0168900208012047} {\emph{ Nucl.\ Instr.\ Meth.} {\bf A 598}  (2009)}
 	\href{http://arxiv.org/abs/0807.2026} {arXiv:0807.2026} 
 	
\bibitem{Raether}
H. Raether, \emph{Elecron avalanches and breakdown in gases}, {Butterworths, Washington, 1964}
 	

\bibitem{Bressan99}
A. Bressan et al., \emph{High rate behavior and discharge limits in micro-pattern detectors},\href{http://www.sciencedirect.com/science/article/pii/S0168900298013175} {\emph{ Nucl.\ Instr.\ Meth.} {\bf A 424}  (1999)}

\bibitem{Bachmann02}
S. Bachmann et al., \emph{High rate behavior and discharge limits in micro-pattern detectors},\href{http://www.sciencedirect.com/science/article/pii/S0168900201009317#} {\emph{ Nucl.\ Instr.\ Meth.} {\bf A 479}  (2002)}
 
\bibitem{Charles11}
G. Charles et al., \emph{Discharge studies in Micromegas detectors in low energy hadron beams},\href{http://www.sciencedirect.com/science/article/pii/S0168900211010217} {\emph{ Nucl.\ Instr.\ Meth.} {\bf A 648}  (2011)}

\bibitem{Arazi12}
L. Arazi et al., \emph{THGEM-based detectors for sampling elements in DHCAL: laboratory and beam evaluation}, \jinst{7}{2012}{C05011}
\href{http://arxiv.org/ftp/arxiv/papers/1112/1112.1915.pdf} {arXiv:1112.1915}
 	
\bibitem{Brau13}
J. Brau, Y. Okada and N. Walker, \emph{The International Linear Collider technical design report},
\href{http://www.linearcollider.org/ILC/Publications/Technical-Design-Report} {TDR} (2013)

\bibitem{Linssen13}
L. Linssen, A. Miyamoto,M. Stanitzki and H. Weerts, \emph{Physics and detectors at CLIC: CLIC conceptual design report},
\href{http://arxiv.org/abs/1202.5940} {arXiv:1202.5940} 

\bibitem{sid09}
H. Aihara, P. Burrows and M. Oreglia, \emph{SiD Letter of Intent},
\href{http://arxiv.org/abs/0911.0006} {arXiv:0911.0006}


\bibitem{Bressler13}
S. Bressler et al., \emph{Beam studies of novel THGEM-based potential sampling elements for Digital Hadron Calorimetry}, \jinst{8}{2013}{P07017}
\href{http://arxiv.org/abs/1305.4657} {arXiv:1305.4657}

\bibitem{Arazi13}
L. Arazi et al., \emph{Beam Studies of the Segmented Resistive WELL: a Potential Thin Sampling Element for Digital Hadron Calorimetry}, {Presented at the $13^{th}$ Vienna Conference on Instrumentation and submitted to its proceedings, February 2013}
\href{http://arxiv.org/abs/1305.1585} {arXiv:1305.1585}

\bibitem{Arazi13_2}
L. Arazi, M. Pitt, S. Bressler, L. Moleri, A. Rubin, S. Sh. Shilstein and A. Breskin, \emph{Laboratory studies of THGEM-based WELL structures with resistive anodes}, {Submitted to JINST}  \href{http://http://arxiv.org/abs/1310.6183} {arXiv:1310.6183}

\bibitem{Weiss57}
J. Weiss and W. Bernstein, \emph{The Current Status of W, the Energy to Produce One Ion Pair in a Gas},\href{http://www.rrjournal.org/doi/abs/10.2307/3570413} {\emph{ Radiation Research} {\bf 7}  (1957)}

\bibitem{Sauli77}
F. Sauli, \emph{Principles of Operation of Multiwire Proportional and Drift Chambers}, {\emph{ CERN 77-09} May (1977)}

\bibitem{Shalem06}
C. Shalem, R. Chechik, A. Breskin and K. Michaeli, \emph{Advances in Thick GEM-like gaseous electron multipliers - Part I: atmospheric pressure operation},\href{http://www.sciencedirect.com/science/article/pii/S016890020502680X} {\emph{ Nucl.\ Instr.\ Meth.} {\bf A 558}  (2006)}

\bibitem{SignalExpress}
\href{http://www.ni.com/labview/signalexpress}{Signal Express web page}

\bibitem{Paschen1889}
F. Paschen, \emph{Ueber die zum Funkenubergang in Luft, Wasserstoff und Kohlensaure bei verschiedenen Drucken erforderliche Potentialdifferenz},
\href{http://onlinelibrary.wiley.com/doi/10.1002/andp.18892730505/abstract}
{\emph{ Wiedemann Annalen der Physik und Chemie} {\bf 37}  (1889)}

\bibitem{Cortesi09}
M. Cortesi et al., \emph{THGEM operation in Ne and Ne/CH$_4$}, \jinst{4}{2009}{P08001}.


\bibitem{Breskin10}
A. Breskin et al., \emph{The THGEM: A thick robust gaseous electron multiplier
for radiation detectors},\href{http://www.sciencedirect.com/science/article/pii/S0168900210004390} {\emph{ Nucl.\ Instr.\ Meth.} {\bf A 623}  (2010)}

\bibitem{Tessarotto10}
F. Tessarotto et al., \emph{Development of THGEM-based photon detectors for Cherenkov Imaging Counters}, 
\jinst{5}{2010}{P03009}.


\bibitem{Azevedo2010}
C.D.R. Azevedo, M. Cortesi, A.V. Lyashenko, A. Breskin, R. Chechik, J. Miyamoto, V. Peskov, J. Escada, J.F.C.A. Veloso, J.M.F. dos Santos. \emph{Towards THGEM UV-photon detectors for RICH: on single-photon detection efficiency in Ne/CH4 and Ne/CF4}, \jinst{5}{2010}{P01002}


\bibitem{Alexeev09}
M. Alexeev et al., \emph{The quest for a third generation of gaseous photon detectors for Cherenkov imaging counters},\href{http://www.sciencedirect.com/science/article/pii/S0168900209010560} {\emph{ Nucl.\ Instr.\ Meth.} {\bf 1 610}  (2009)}
 	\href{http://arxiv.org/pdf/0907.3577.pdf} {arXiv:0907.3577}  
 

\bibitem{Rubin_opt13}
A. Rubin et al., \emph{Optical readout: a tool for studying gas-avalanche processes}, 
\jinst{8}{2013}{P08001}.
\href{http://arxiv.org/abs/1305.1196} {arXiv:1305.1196}  

\end{thebibliography}
\end{document}